\documentclass[CRPHYS,Unicode,manuscript]{cedram}

\usepackage{amsmath,amssymb}
\usepackage{caption}

\usepackage{miller}

\usepackage[sort]{cite}

\newcommand{\ie}{\textit{i.e.}}
\newcommand{\cf}{\textit{cf}}
\newcommand{\abinitio}{\textit{ab initio}}
\newcommand{\Abinitio}{\textit{Ab initio}}
\newcommand{\insitu}{\textit{in situ}}
\newcommand{\vasp}{\textsc{Vasp}}
\newcommand{\pwscf}{\textsc{Pwscf}}

\newcommand{\barT}[1]{\bar{\bar{#1}}}
\newcommand{\ud}{\mathrm{d}}
\DeclareMathOperator{\diag}{diag}

\usepackage{xfrac}

\newcommand{\be}{\begin{equation}} 
\newcommand{\ee}{\end{equation}}
\newcommand{\bn}{\begin{eqnarray}} 
\newcommand{\en}{\end{eqnarray}}

\title{Screw dislocations in BCC transition metals:
	from ab initio modeling to yield criterion}

\author[E. Clouet]{\firstname{Emmanuel} \lastname{Clouet}\IsCorresp}
\address{Université Paris-Saclay, CEA, Service de Recherches de Métallurgie Physique, 91191, Gif-sur-Yvette, France}
\email[EC]{emmanuel.clouet@cea.fr}

\author[B. Bienvenu]{\firstname{Baptiste} \lastname{Bienvenu}}
\address{Université Paris-Saclay, CEA, Service de Recherches de Métallurgie Physique, 91191, Gif-sur-Yvette, France}

\author[L. Dezerald]{\firstname{Lucile} \lastname{Dezerald}}
\address{Institut Jean Lamour, CNRS UMR 7198, Université de Lorraine, F-54000 Nancy, France}

\author[D. Rodney]{\firstname{David} \lastname{Rodney}}
\address{Institut Lumière Matière, Université Lyon 1 - CNRS, Villeurbanne F-69622, France}


\begin{abstract} 
We show here how density functional theory calculations can be used to predict the temperature- and orientation-dependence of the yield stress of body-centered cubic (BCC) metals in the thermally-activated regime where plasticity is governed by the glide of screw dislocations with a $1/2\,\hkl<111>$ Burgers vector. Our numerical model incorporates non-Schmid effects, both the twinning/antitwinning asymmetry and non-glide effects, characterized through \abinitio{} calculations on straight dislocations. The model  uses the stress-dependence of the kink-pair nucleation enthalpy predicted by a line tension model also fully parameterized on \abinitio{} calculations. The methodology is illustrated here on BCC tungsten but is applicable to all BCC metals.  Comparison with experimental data allows to highlight both the successes and remaining limitations of our modeling approach.
\end{abstract}

\begin{document}
\maketitle

\section{Introduction}

The plasticity of body-centered cubic (BCC) metals has attracted and will continue to attract a lot of attention for both technological and scientific reasons. Technologically, BCC metals are ubiquitous among structural materials due to their high yield strength and toughness \cite{Ashby2018}. For example, mild steels have a BCC matrix of $\alpha$-Fe hardened by various types of alloying elements, precipitates and second-phase particles. Another example is tungsten, which is foreseen to be used in fusion reactors due to its high density and elevated melting point \cite{Rieth2013}.

Plasticity in BCC metals is mainly due to the glide of screw dislocations with a $1/2\,\langle 111 \rangle$ Burgers vector \cite{Hirsch1960}. These dislocations have proved to exhibit unusual properties, which translate directly in surprising features of BCC plasticity at the macroscopic scale \cite{Christian83}. Based on atomistic simulations \cite{Vitek1970} and \insitu{} transmission electron microscopy \cite{Louchet1979,Caillard2010}, we know that screw dislocations feel a strong lattice resistance and glide through a thermally-activated process, which involves the nucleation and propagation of kink-pairs along the dislocation line. Different glide planes, $\{ 110\}$, $\{ 121\}$ \cite{Argon1966,Spitzig1970}, even $\{ 123\}$ \cite{Caillard2018}, have been observed. At low temperature in most metals, $\{ 110\}$ planes dominate but it remains unclear to this date whether glide in different planes results from different glide mechanisms or from a combination of elementary glide events in different $\{ 110\}$ planes. Very surprising glide sequences in different $\{ 121\}$ planes have also been recently observed in tungsten by \insitu{} TEM \cite{Caillard2018}. The thermal activation of the screw dislocation mobility implies a rapid increase of the yield stress at low temperatures \cite{Caillard2003}. However, a still much debated observation is that the best prediction of the 0\,K limit of the yield stress, the so-called Peierls stress, based on atomistic models is two to three times larger than experimental extrapolations \cite{Groger2007,Proville2012,Freitas2018}. Another surprising feature of the screw dislocations is that they do not obey the classical Schmid law \cite{Schmid1924}, which states that dislocation motion is driven only by the resolved shear stress, \ie{} the part of the stress tensor, which produces a Peach-Koehler force in the glide plane of the dislocation. Screw dislocations do not follow Schmid law for two reasons \cite{Duesbery1998}. First, due to the asymmetry of the BCC lattice, positive and negative shear stresses on planes other than $\{110\}$ are not equivalent, resulting in the so-called twinning/antitwinning (T/AT) asymmetry. Second, the screw dislocations are affected not only by the resolved shear stress but also by components of the stress tensor which do not produce a Peach-Koehler force, inducing so-called non-glide effects.

The lattice resistance as well as non-Schmid effects result from the structure of the BCC lattice and the core structure of the $1/2\,\langle 111 \rangle$ screw dislocation. Both effects can only be accounted for through atomistic simulations. Empirical potentials have been used since the 1970s \cite{Vitek1970} and have led to valuable yet sometimes inconsistent results between potentials or even with experimental evidence. In particular, most interatomic potentials predict a three-fold degenerate core, which glides in $\{ 121\}$ planes, while as mentioned above, slip occurs experimentally at low temperature mostly in $\{ 110\}$ planes. However, with the increase of computing power, it has become possible at the turn of the 21st century \cite{Ismail2000} to model screw dislocations using \abinitio{} density functional theory (DFT) calculations. These calculations have led to a wealth of new results, showing that in all pure BCC transition metals, the dislocation core is not degenerate nor asymmetrically extended, but is rather non-degenerate and compact \cite{Ismail2000,Woodward2002,Frederiksen2003,Ventelon2007}. DFT has also allowed to study the T/AT asymmetry \cite{Dezerald2016} and non-glide effects \cite{Kraych2019} on perfectly straight screw dislocations, as will be detailed below. At finite temperatures, dislocation glide involves kink pairs, which extend over several tens of Burgers vectors along the dislocations, requiring simulation cells too large for \abinitio~calculations. In this case, classical molecular dynamics simulations with carefully tested interatomic potentials remain highly valuable \cite{Domain2005,Chaussidon2006,Gilbert2011,Po2016}. Another approach is to introduce a coarse-grained model adjusted on DFT data, for instance a line tension model \cite{Itakura2012,Proville2013,Dezerald2015,He2019}, as described below. To study collective effects, one then needs to go one scale up and use for instance dislocation dynamics \cite{Chaussidon2008,Po2016}. However, in BCC metals at low temperature, collective effects are not dominant, in contrast with face-centered-cubic (FCC) metals and the yield strength can be directly predicted from the behavior of an isolated screw dislocation \cite{Groger08b}.

In the past 15 years, important results on the mobility of screw dislocations in BCC metals have been obtained at different scales but in an uncoordinated way, using data from different sources and obtained with different codes. In the present paper, we would like to show on the specific example of tungsten how data obtained from \abinitio{} calculations can be used in a consistent way to develop a yield criterion including non-Schmid effects able to predict the plastic strength as a function of both the temperature and the direction and sign of the deformation axis.

\section{Dislocations and ab initio calculations}
\label{sec:methods}

\Abinitio{} calculations of dislocations rely on the density functional theory (DFT)
\cite{Hohenberg1964,Kohn1965}.  
Such an approach is computationally demanding and can handle only small systems, 
typically a few hundred atoms, at most a few thousands on supercomputers. 
As a consequence, \abinitio{} calculations have mainly been restricted until now to model infinite straight dislocations, 
in order to minimize the length of the simulation cell in the direction of the dislocation line. 
In the other directions, it is necessary to carefully handle 
the elastic strain field created by the dislocation, as this elastic field leads to a displacement
which varies as the logarithm of the distance to the dislocation line
and therefore does not vanish at short range.
In addition, is it not possible, for topological reasons, to model a single dislocation in a supercell with full 
periodic boundary conditions, since the displacement discontinuity created by the dislocation
needs to be closed by another defect.
Several approaches specific to dislocations have been developed.
We only give below a brief overview of these approaches which can be divided in two categories
and refer to recent review articles \cite{Woodward2005,Rodney2017,Clouet2018h}
for technical details.

\begin{figure}[btp]
	\begin{center}
		\includegraphics[width=0.9\linewidth]{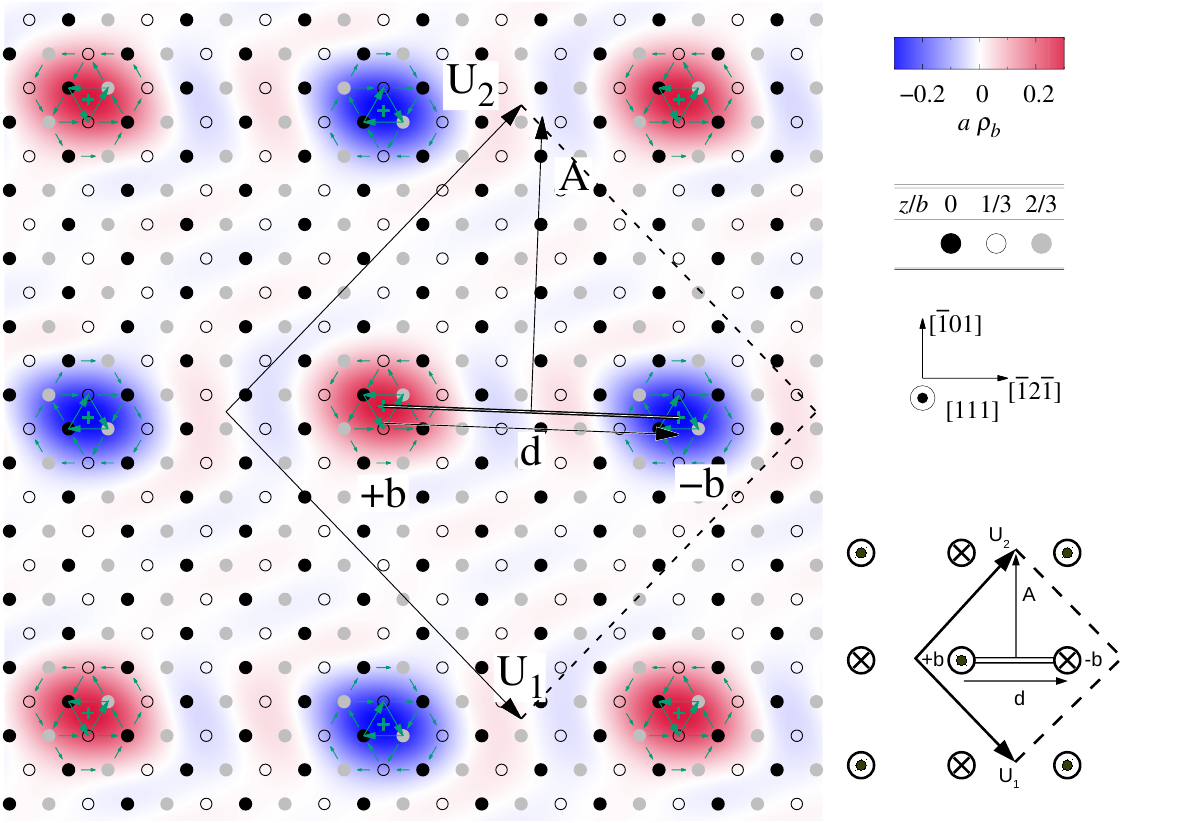}
	\end{center}
	\caption{Typical simulation cell used to model $1/2\,\hkl<111>$ screw dislocations
	in BCC metals. The cell contains a dislocation dipole, which forms a quadrupolar arrangement of dislocations of opposite Burgers vectors through the periodic boundary conditions applied in all directions.
	The dipole is defined by its Burgers vector $\vec{b}$,
	the dipole vector $\vec{d}$ joining both dislocation lines,
	and the cut vector $\vec{A}$, 
	with the corresponding discontinuity surface indicated by a double black line. 
	$\vec{U}_1$ and $\vec{U}_2$ are the periodicity vectors of the simulation cell perpendicular
	to the dislocation line. 
	The atomic structure of the dislocations is shown with differential displacement and Nye tensor maps.
	In this projection perpendicular to the dislocation line,
	atoms are shown as circles with a color depending on their \hkl(111) plane 
	in the original perfect lattice.
	Arrows between atomic columns are proportional to the differential displacement 
	created by the dislocation in the direction of the Burgers vector.
	The color map shows the dislocation density $\rho_b$ normalized by the lattice parameter $a$.
	(Figure adapted from \cite{Clouet2018h})
	}
	\label{fig:PBC_sketch}
\end{figure}

\paragraph{Cluster approach:}
	one can embed a single dislocation in an infinite cylinder with its axis 
	along the line defect.  Atoms in the outer surface of the cylinder
	are either kept fixed at their positions predicted by elasticity theory
	or relaxed according to the harmonic response of the crystal 
	\cite{Sinclair1978,Woodward2002,Woodward2005}.  
	Only the atoms inside the cylinder, \ie{} the atoms close to the dislocation cores,
	are relaxed according to the Hellman-Feynman forces calculated \abinitio{}.
	With such boundary conditions, one truly models a single dislocation 
	in an infinite crystal. 
	The main drawback of the approach is the difficulty to isolate 
	the dislocation contribution to the calculated excess energy. 
	One misses in \abinitio{} calculations a rigorous local projection of the energy which would allow
	to separate the excess part coming from the dislocation 
	and the one caused by the boundary, \ie{} the external surface 
	of the cylinder.

\paragraph{Dipole approach:}
	to keep full periodic boundary conditions, and thus avoid the need for an external surface,
	one has to introduce a dislocation dipole in the simulation cell,
	with two dislocations of opposite Burgers vectors. 
	One then models an infinite periodic array of dislocations.
	The excess energy is then due to the core energy of the dislocations as well as their mutual elastic interactions, which involve the dislocations inside the supercell as well as their periodic images. The elastic energy can be evaluated quantitatively 
	using anisotropic elasticity \cite{Clouet2018h}, yielding after subtraction the energy of a single dislocation, \ie{}
	its core energy and variation with the dislocation position in the atomic lattice.
	Among the different periodic arrangements proposed to model dislocation dipoles,
	quadrupolar arrangements have to be preferred, as such arrangements minimize
	the elastic interaction between the dislocations of the dipole and with their periodic images. This arrangement leads to well-converged dislocation properties
	for small simulation cells compatible with \abinitio{} calculations, as long as the dislocation core is compact \cite{Clouet2009}. 
	In BCC metals for instance, it is sufficient to use a supercell containing 135 atoms
	per $b$ ($b$ is the norm of the Burgers vector)
	to model $1/2\,\hkl<111>$ screw dislocations. 
	Such supercell containing a dislocation dipole is shown in Fig. \ref{fig:PBC_sketch}. 

	Another advantage of full periodic boundary conditions is that one can relate
	the stress variations observed in the supercell to the dislocation relative positions.
	Noting $\vec{d}$ the vector joining the center of the $+\vec{b}$ dislocation
	to the $-\vec{b}$ dislocation, one defines the cut vector of the dipole as 
	$\vec{A}=\vec{l}\wedge\vec{d}$, where $\vec{l}$ is the dislocation line vector
	(see Fig. \ref{fig:PBC_sketch} for an example).
	According to linear elasticity theory, the energy variation caused by a homogeneous strain 
	$\barT{\varepsilon}$ is\footnote{Here, and in the following equations, 
	we use Einstein summation convention on repeated indexes.} 
	\begin{equation}
		\Delta E(\barT{\varepsilon}) = \frac{1}{2} S \, C_{ijkl} \, \varepsilon_{ij} \, \varepsilon_{kl}
		+ C_{ijkl} \, b_i \, A_j \, \varepsilon_{kl},
		\label{eq:dE_epsi}
	\end{equation}
	where $C_{ijkl}$ are the elastic constants of the perfect crystal
	and $S$ is the area of the supercell perpendicular 
	to the dislocations line vector $\vec{l}$.
	This energy is defined per unit length of the supercell in the $\vec{l}$ direction.
	The stress is then simply obtained by computing the derivative 
	of the energy, leading to
	\begin{equation}
		\sigma_{ij}(\barT{\varepsilon}) = \frac{1}{S} \frac{\partial \Delta E}{\partial \varepsilon_{ij}}
		= C_{ijkl} \left( \varepsilon_{kl} - \varepsilon_{kl}^0 \right),
		\label{eq:stress_epsi}
	\end{equation}
	with the plastic strain defined as
	\begin{equation}
		\varepsilon_{kl}^0 = - \frac{b_k \, A_l + b_l \, A_k}{2S}.
		\label{eq:plastic_strain}
	\end{equation}
	According to Eq. \ref{eq:stress_epsi}, a homogeneous strain $\barT{\varepsilon}$,
	equal to the plastic strain $\barT{\varepsilon}^0$ introduced in the supercell 
	when creating the dislocation dipole, needs to be applied to the simulation cell
	to maintain a zero stress \cite{Cai2003,Daw2006}.
	A different homogeneous strain can be applied to reach another target stress.
	Eqs. \ref{eq:stress_epsi} and \ref{eq:plastic_strain} also show that if the relative positions of both dislocations vary, \ie{} if the cut-vector $\vec{A}$ varies, a stress will build up in the supercell. 
	The stress variation obtained by \abinitio{} calculations through
	generalized Hellman-Feynman forces can thus be used to compute the dislocation
	relative positions.
	In this way, one can extract dislocation trajectories from \abinitio{}
	energy barriers calculations \cite{Chaari2014,Dezerald2016}, as done in section \ref{sec:PeierlsPotential}.

All results described in this article for the $1/2\,\hkl<111>$ screw dislocation in tungsten
were obtained with the quadrupolar setup of Fig. \ref{fig:PBC_sketch} in a periodic supercell
containing 135 atoms.
\Abinitio{} calculations were performed with either the \pwscf{} or \vasp{} codes,
as reported in the original works 
\cite{Dezerald2014,Dezerald2015,Dezerald2016,Kraych2019} along with the corresponding DFT parameters. 
Unpublished results appearing below were obtained 
with the calculations parameters given in Appendix \ref{app:abinitio}.

\section{Dislocation core structures and energy landscape}
\label{sec:landscape}

As first realized by Edagawa \emph{et al.} \cite{Edagawa97a,Edagawa97b}, the $1/2\,\hkl<111>$ screw dislocation can in principle take any position in the $\{111\}$ plane perpendicular to its line. Its core energy will depend on the position, yielding a two-dimensional (2D) energy landscape, which will show low energy regions near stable positions, separated by energy barriers that dominate the low-temperature glide of the dislocation as well as higher energy regions.

\begin{figure}[tbh]
	\begin{center}
	\includegraphics[width=0.9\linewidth]{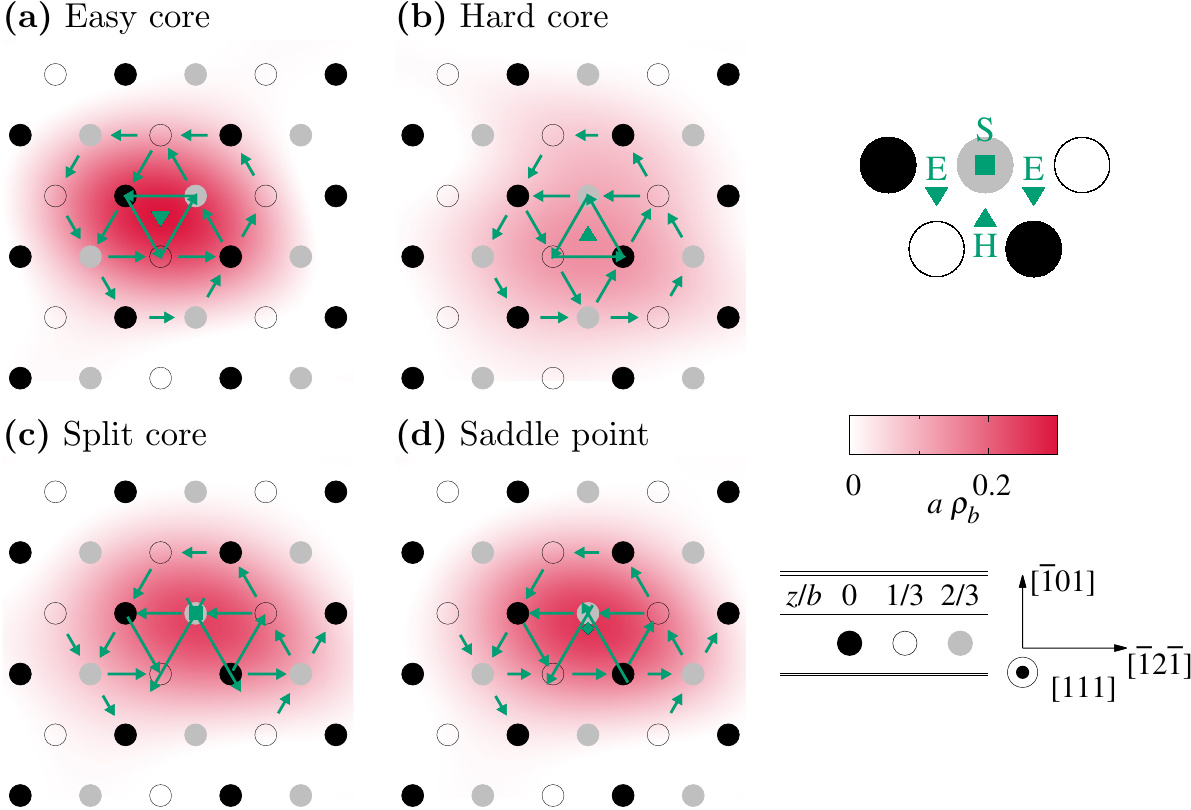}
	\end{center}
	\caption{Core structures of a screw dislocation in BCC tungsten \cite{Dezerald2014,Dezerald2016}: (a) easy, (b) hard, (c) split cores,
	and (d) saddle configuration.
	The centers of the dislocation are indicated
	by colored symbols and are sketched in the right panel. Differential displacements and Nye tensors are shown as in Fig. \ref{fig:PBC_sketch}.
	}
	\label{fig:core}
\end{figure}
In order to compute the 2D energy landscape from DFT, we start by considering the high-symmetry positions when the dislocation is located at the center of a triangle of $\langle 111 \rangle$ atomic columns. As illustrated in Fig. \ref{fig:core}, atoms in these columns have different heights and form clockwise helices in upward triangles and anticlockwise helices in downward triangles. Inserting a screw dislocation at the center of such triangle adds a helical displacement field, which either inverses the chirality of the helix or brings all three columns at the same height. The first case, shown in Fig. \ref{fig:core}(a), results in the lowest-energy core configuration of the screw dislocation, called an easy core. The second case, shown in Fig. \ref{fig:core}(b), is by symmetry an energy maximum, called a hard core. We see from the differential displacements and Nye tensor in Fig. \ref{fig:core}(a) that the easy core is compact and symmetrical, as observed in other BCC pure metals \cite{Ismail2000,Woodward2002,Frederiksen2003,Weinberger2013,Dezerald2014} and in contrast with the predictions of many empirical interatomic potentials. A third high-symmetry position is when the dislocation core is located in the immediate vicinity of an atomic column. This configuration shown in Fig. \ref{fig:core}(c) is called a split core \cite{Takeuchi1979}. It does not preserve the 3-fold symmetry and has thus three variants depending on the region from which the dislocation approaches the atomic column.

\begin{figure}[tbh]
	\begin{center}
		\includegraphics[width=0.9\linewidth]{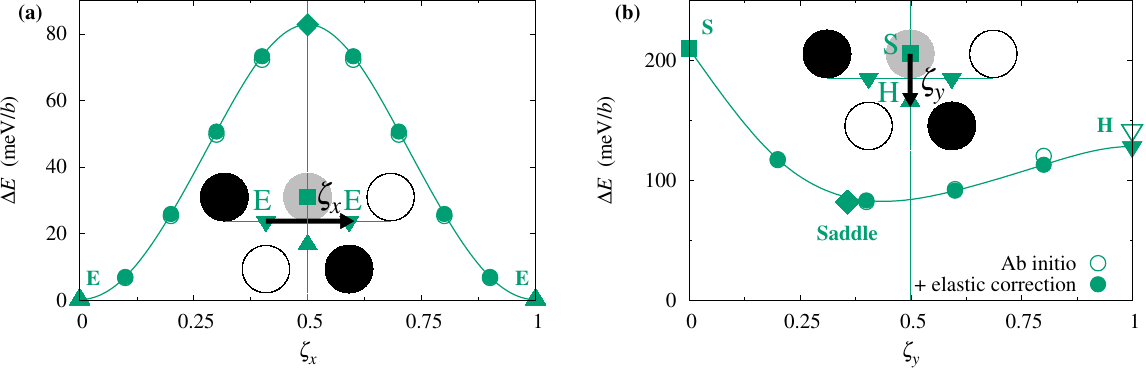}
	\end{center}
	\caption{Energy variation of the dislocation along different paths in BCC tungsten \cite{Dezerald2014}:
	(a) Peierls barrier, \ie{} minimum energy path between easy cores;
	(b) Straight path between split and hard cores. 
	The saddle configuration in (a) is also indicated by a diamond in (b).
	The insets show the path directions
	with the corresponding reaction coordinates, $\zeta_x$ and $\zeta_y$.}
	\label{fig:PeierlsCuts}
\end{figure}

At rest, the screw dislocation is in its low-energy, stable, easy core configuration and glides through a thermally-activated jump to a nearby easy core. As shown in Fig. \ref{fig:PeierlsCuts}(a), we computed the corresponding energy barrier \cite{Dezerald2014}, the so-called Peierls barrier, using the nudged elastic band (NEB) method \cite{Henkelman2000}. 
In the energy barrier calculation, both dislocations composing the dipole are displaced in the same direction
to keep their distance constant.  
As a consequence, the elastic energy along the migration path is almost constant: 
only a small variation arises from a deviation of the dislocation trajectory from a straight path, 
this deviation being in opposite direction for the $+\vec{b}$ and $-\vec{b}$ dislocations.
The very small elastic correction is illustrated in Fig. \ref{fig:PeierlsCuts}(a) by the difference between the open symbols obtained directly from the DFT calculation and the full symbols calculated after subtracting the elastic energy. We see in Fig. \ref{fig:PeierlsCuts}(a) that the energy barrier has a single hump and that the dislocation passes by a saddle configuration at mid-distance between easy cores. The corresponding atomic configuration is shown in Fig. \ref{fig:core}(d). 

In order to obtain 2D information, we computed in the same cell the energy path between the split and hard cores \cite{Dezerald2014}, \ie{} along the ridge which separates the basins of attraction of two successive easy cores. The result is shown in Fig. \ref{fig:PeierlsCuts}(b), where we see a slightly larger, but still rather small, elastic correction. For these calculations, we constrained the difference in altitude between the two atomic columns on either side of the ridge (the white and black columns in the inset of Fig. \ref{fig:PeierlsCuts}(b)) to fix the core position and forbid the dislocation to relax to an easy core configuration during minimization. We find as expected that the hard core is an energy maximum. Less expected, at least from empirical interatomic potential calculations, is that the split core is also an energy maximum, with an energy even higher than the hard core. In-between, there is a minimum, which corresponds to the saddle configuration between easy cores.

\begin{figure}[tb]
	\begin{center}
		\includegraphics[width=0.50\linewidth]{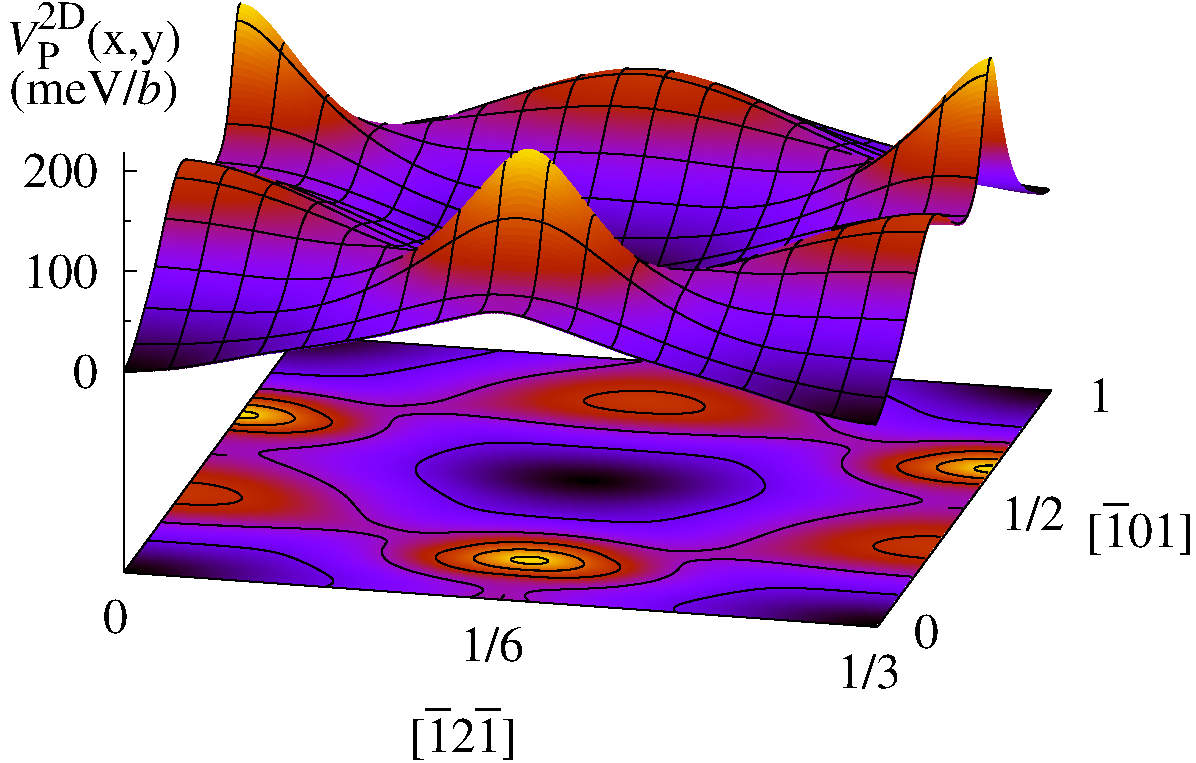}
		\includegraphics[width=0.48\linewidth]{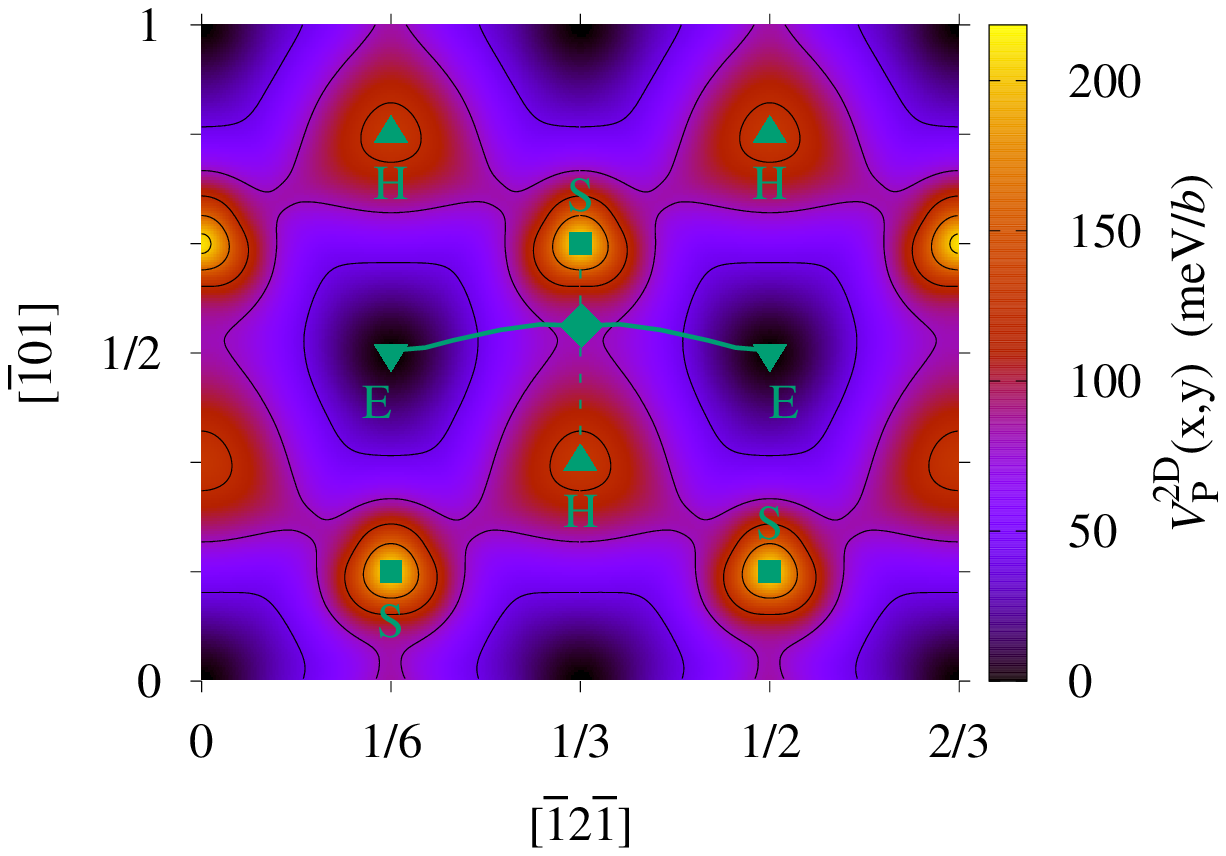}
	\end{center}
	\caption{2D Peierls potential of the $1/2\,\hkl<111>$ screw dislocation in BCC tungsten \cite{Dezerald2014}.
	In the projection on the right, the dislocation trajectory corresponding to the minimum energy path between 
	two easy cores (Fig. \ref{fig:PeierlsCuts}a) is shown with a green solid line,
	while the dashed green line links the split and the hard cores (energy variation shown on Fig. \ref{fig:PeierlsCuts}b).
	The green diamond corresponds to the saddle configuration.
	}
	\label{fig:Peierls2D}
\end{figure}

We constructed a continuous 2D energy landscape, $V_{\rm P}^{\rm 2D}(x,y)$ shown in Fig. \ref{fig:Peierls2D} from the energy along both the Peierls barrier and the split-to-hard line using a Fourier decomposition satisfying the three-fold symmetry of the lattice \cite{Dezerald2014}. For this, the position of the dislocation along the paths was determined by fitting the relaxed atomic positions given by the \abinitio{} calculations to the positions predicted by anisotropic elasticity theory,
using the dislocation position as fitting parameter. 
Similar dislocation trajectories are obtained using the stress variation along the paths mentioned in section \ref{sec:methods}.
We see in Fig. \ref{fig:Peierls2D}(b) that the path between easy cores is not straight, but deviates towards the split core. We will link this deviation to the T/AT asymmetry in section \ref{sec:trajectory}. For now, we note that the 2D landscape highlights that the $\{111\}$ plane is made of energy minima which correspond to the easy core positions, of primary maxima at the split cores and secondary maxima at the hard cores. Similar 2D energy landscapes were obtained in other BCC transition metals, 
with the relative heights of the maxima depending on the element \cite{Dezerald2014}. 
Only Fe differs from the general behavior with an almost constant split-to-hard energy profile near the hard core position. 
Instead of a local maximum, the hard core in Fe is a saddle point which links three different ground states.

\section{From the energy landscape to the Peierls stress}
\label{sec:PeierlsPotential}

\subsection{Peierls enthalpy barrier and Peierls stress}
\label{sec:PeierlsStress}

A central quantity to characterize the mobility of a dislocation is its Peierls stress $\tau_{\rm P}$, \ie{} the critical resolved shear stress to apply in order to induce motion of the dislocation at zero Kelvin. The Peierls stress can be determined using quasi-static calculations where an increasing shear stress is applied to the simulation cell in increments followed by energy minimizations until the dislocation starts to move. This method however requires a deep relaxation of the interatomic forces, which costs high CPU time with \abinitio{} calculations. Another method to determine the Peierls stress is to compute the Peierls barrier discussed above under an applied stress. The Peierls stress then corresponds to the applied stress at which the maximum of the Peierls barrier disappears. This method is more accurate than quasi-static calculations because it relies on the convergence of the energy, which is more easily achieved than forces within \abinitio{} calculations. We will use this method in the following.

\begin{figure}[!bt]
	\begin{center}
        \includegraphics[clip, width=0.9\linewidth]{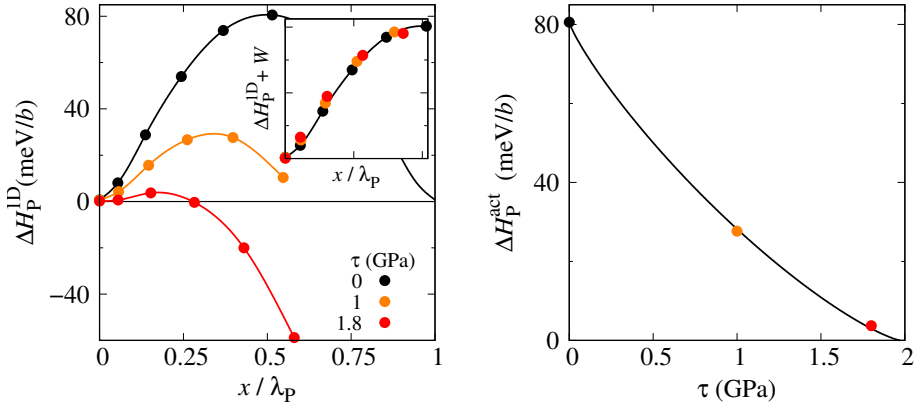}
	\end{center}
	\caption{
	(a) Peierls enthalpy barriers calculated by DFT in BCC tungsten for different applied resolved shear stresses $\tau$ as a function of the dislocation position, $x$, between Peierls valleys ($\lambda_P$ is the distance between Peierls valleys). 
	The inset shows the corrected barriers $\Delta H_ {\rm P}^{\rm 1D}(x)+W(x)$, where $W(x)=\tau\,b\,x$ is the work of the applied stress, for the different stresses (symbols) and the interpolation obtained without stress (line). 
	(b) Maximum Peierls enthalpy as a function of  applied resolved shear stress, with a power-law fit.
	}
	\label{fig:PeierlsEnthalpy}
\end{figure}

Peierls barriers under stress were computed by applying a strain tensor to the simulation cell to produce the targeted shear stress through Eq. \ref{eq:stress_epsi} followed by a NEB calculation~\cite{Dezerald2016}. Examples for a shear stress resolved in the $(\bar{1}01)$ glide plane and along the $[111]$ Burgers vector of a screw dislocation in tungsten are presented in Fig.~\ref{fig:PeierlsEnthalpy}~(a). We note that to reduce the computed cost, we performed the NEB calculations only on the first half of the paths. We see that the maximum of the energy barrier decreases under increasing shear stress, as expected. Although the calculations are done under a constant strain, if the plastic strain generated along the path is small enough to assume that the stress is constant to first order, the energies obtained here correspond to the enthalpy under the targeted stress. To evaluate the Peierls stress, we extracted from Fig.~\ref{fig:PeierlsEnthalpy}~(a) the Peierls enthalpies $\Delta H_ {\rm P}^{\rm act}(\tau)$, \ie{} the maximum of the enthalpy profiles, which we plotted as a function of the applied shear stress in Fig.~\ref{fig:PeierlsEnthalpy}~(b). The value of $\tau_{\rm P}$ is then determined by extrapolation using a power law interpolation to find the resolved shear stress $\tau_{\rm P}$ satisfying $\Delta H_ {\rm P}^{\rm act}(\tau_{\rm P})=0$.
The fit, displayed as a solid black line in Fig.~\ref{fig:PeierlsEnthalpy}~(b), yields $\tau_{\rm P} = 1970$~MPa.

We note that the theoretical value of the Peierls stress is significantly larger than the value extrapolated from experiments (900 MPa \cite{Dezerald2015}). Similar discrepancies between theoretical and experimental values of the Peierls stress have been reported in all BCC metals using different atomistic models, from simple pair potentials \cite{Suzuki1970,Basinski1971}, to more advanced embedded atom potentials \cite{Chaussidon2006}, to \abinitio{} calculations as performed here \cite{Woodward2001,Woodward2002,Romaner2010,Ventelon2013,Dezerald2014,Dezerald2016}. Potential collective \cite{Bulatov2002,Groger2007} and quantum \cite{Suzuki1970,Proville2012,Barvinschi2014} origins have been proposed but are still a matter of debate \cite{Freitas2018}.

In the following section, we model the effect of the stress on the 2D energy landscape of the dislocation by assuming that the stress only tilts the potential through the plastic work ($\tau b x$ per unit length of the dislocation if $\tau$ is the $\hkl<111> \hkl(-101)$ resolved shear stress), but does not affect the potential itself. In the inset of Fig. \ref{fig:PeierlsEnthalpy}(a), we show the corrected energy $\Delta H_{\rm P}^{\rm 1D}(x) + \tau b x$, which is indeed independent of the applied stress. We note that the measured variation of the Peierls potential under stress is highly sensitive to the definition used for the dislocation position $x$: it is found independent only when the position is derived from the stress variation observed along the path (\cf{} methods in section \ref{sec:methods}). Similar independence was observed in other BCC metals \cite{Dezerald2016,Kraych2019}, contrasting with an earlier observation on a specific dislocation in aluminum modeled with an interatomic potential \cite{Rodney2009}.

\subsection{Deviation to the Schmid law and dislocation trajectory}
\label{sec:trajectory}

\begin{figure}[!b]
	\begin{center}
		\includegraphics[width=0.49\linewidth]{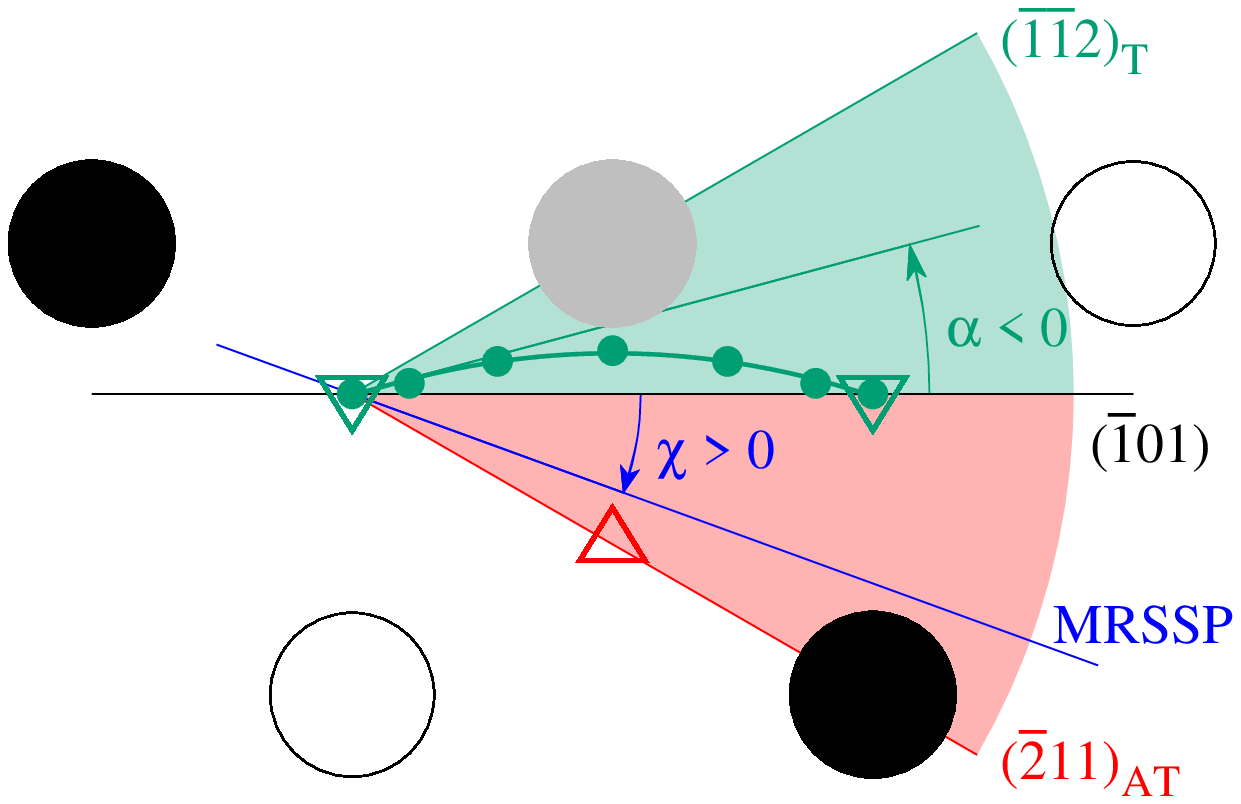}
		\hfill
		\includegraphics[width=0.49\linewidth]{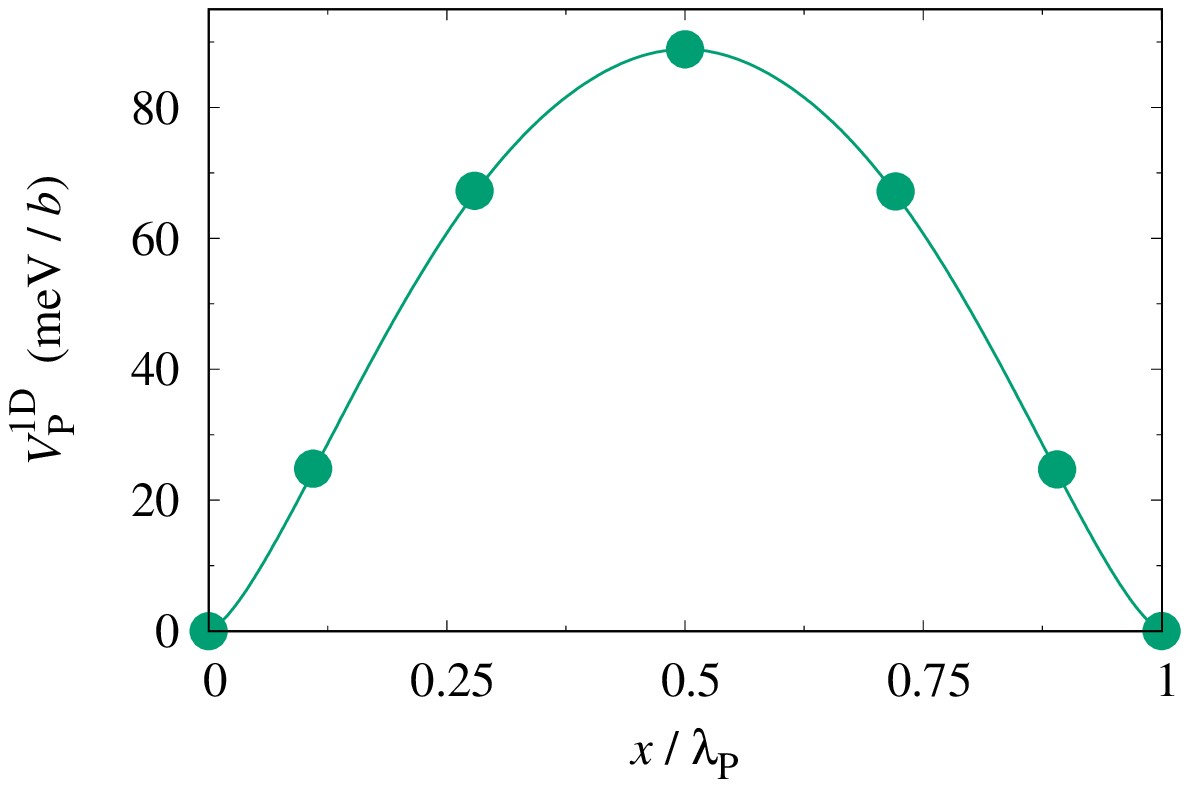}
	\end{center}
	\caption{Trajectory of a $1/2\,\hkl[111]$ screw dislocation between two neighboring easy core positions in BCC tungsten,
	with corresponding Peierls potential\textsuperscript{a}. 
	The dislocation position was deduced from the \abinitio{} stress variation along the minimum energy path (Eqs. \ref{eq:stress_epsi} and \ref{eq:plastic_strain}). 
	The angle $\alpha$ defines the orientation of the tangent to the trajectory in the first half of the path,
	while the maximum resolved shear stress plane (MRSSP) is defined by the angle $\chi$. Downward and upward triangles indicate easy and hard core positions, respectively.
	\\
	\footnotesize \textsuperscript{a}The Peierls potential shown here and further used in the article 
	slightly differs from the one published in Ref. \cite{Dezerald2016} and shown in Fig. \ref{fig:PeierlsEnthalpy} 
	because of small differences in the \abinitio{} parameters and dislocation setup 
	(\cf{} appendix \ref{app:abinitio} for a description of the parameters used in the present work).
	}
	\label{fig:trajectory}
\end{figure}

The calculations above were performed with an applied shear stress resolved in the $(\bar{1}01)$ glide plane of the screw dislocation. With the notations of Fig. \ref{fig:trajectory}, this corresponds to a $\chi=0$ maximum resolved shear stress plane (MRSSP). Similar calculations can be done with $\chi \neq 0$ by adjusting the applied strain tensor. If Schmid's law applies, \ie{} if dislocation glide is activated 
when the shear stress resolved in the glide plane reaches the Peierls stress in this plane, we expect the yield stress to follow the relation
\be
\tau_{\rm P}(\chi) = \frac{\tau_{\rm P}^0}{\cos(\chi)},
\ee
where $\tau_{\rm P}^0=\tau_{\rm P}(\chi=0)$ is the Peierls stress in the $(\bar{1}01)$ glide plane computed above. The above yield criterion is symmetrical between $\chi >0$ and $\chi < 0$, a characteristic of Schmid's law. However, as mentioned in Introduction, a hallmark of low-temperature plasticity in BCC metals is that they do not obey Schmid's law. In particular, the yield stress is lower when the MRSSP is on one side of the glide plane, the twinning region which corresponds to $\chi <0$ with the notations of Fig. \ref{fig:trajectory}, than on the other side, the antitwinning region with $\chi >0$. This twinning/antitwinning (T/AT) asymmetry is a consequence of the lack of symmetry of the BCC lattice with respect to $\hkl{110}$ planes. This lack of symmetry is also at the origin of the deviation of the dislocation trajectory, which was visible in Fig. \ref{fig:Peierls2D} and is reproduced with more details in Fig. \ref{fig:trajectory}, using now the dislocation position defined from the stress variation along the path (Eqs. \ref{eq:stress_epsi} and \ref{eq:plastic_strain}). We see in this figure that the trajectory is deviated towards the split core (i.e. the atomic column in gray), which crystallographically always lies in the twinning region, and away from the hard core (upward red triangle), which necessarily lies in the antitwinning region. This is a direct consequence of the position of the saddle configuration along the split-to-hard line in Fig. \ref{fig:PeierlsCuts}, which is closer to the split core than to the hard core. Below we show that we can quantitatively relate the amplitude of the deviation of the trajectory to the amplitude of the T/AT asymmetry~\cite{Dezerald2016}.

As mentioned above, we assume that the effect of an applied stress on the 2D energy landscape discussed in section \ref{sec:landscape} is to add a linear contribution corresponding to the plastic work, yielding an enthalpy variation
\begin{equation}
	\Delta H^{\rm 2D}_{\rm P}(x,y) = V_{\rm P}^{\rm 2D}(x,y) - \tau_{yz}\,b\,x + \tau_{xz}\,b\,y,
	\label{eq:Peierls2D}
\end{equation}
with $x$ and $y$, the coordinates of the dislocation respectively in the $\hkl[-12-1]$ glide direction and $\hkl[-101]$ perpendicular direction and $\tau_{xz}$ and $\tau_{yz}$, the two components of the stress tensor which produce a Peach-Koehler force on the dislocation.
Considering a resolved shear stress $\tau$ applied in a MRSSP making an angle $\chi$ with respect to the $\hkl(-101)$ glide plane,
both components are written as
$\tau_{xz}=-\tau\sin{(\chi)}$ and $\tau_{yz}=\tau\cos{(\chi)}$, and the enthalpy variation becomes
\begin{equation*}
	\Delta H^{\rm 2D}_{\rm P}(x,y) = V_{\rm P}^{\rm 2D}(x,y) - \tau\,b \left[ x \cos{(\chi)} + y \sin{(\chi)} \right].
\end{equation*}
The saddle point value of this 2D function between two easy cores is the enthalpy barrier $\Delta H_{\rm P}^{\rm act}(\tau,\chi)$ opposing dislocation glide, which was shown in Fig. \ref{fig:PeierlsEnthalpy}~(b) for $\chi=0$.
The Peierls stress $\tau_{\rm P}(\chi)$ is then defined as the minimal applied stress $\tau$ 
for which this enthalpy barrier vanishes.

\Abinitio{} calculations have shown that the trajectory between easy cores (Fig. \ref{fig:Peierls2D})
is not sensitive to the applied shear stress $\tau$ nor to the other stress components 
\cite{Dezerald2016,Kraych2019}.
Assuming that this trajectory does not vary, one can define a 1D functional for the enthalpy barrier between Peierls valleys
\begin{equation}
	\begin{split}
	\Delta H^{\rm 1D}_{\rm P}(x) &= \Delta H^{\rm 2D}_{\rm P}(x,\bar{y}(x)) \\
		&= V_{\rm P}^{\rm 1D}(x) - \tau\,b \left[ x \cos{(\chi)} + \bar{y}(x) \sin{(\chi)} \right],
	\end{split}
	\label{eq:Peierls1D}
\end{equation}
where $\bar{y}(x)$ is the dislocation trajectory and $V_{\rm P}^{\rm 1D}(x) = V_{\rm P}^{\rm 2D}[x,\bar{y}(x)]$ is the energy barrier, shown in Fig. \ref{fig:PeierlsEnthalpy} for $\tau=0$.
At the Peierls stress, there exists an unstable position $x^*$ 
where the first and second derivatives of the enthalpy are null, leading to
\begin{align}
	\left. \frac{\partial V_{\rm P}^{\rm 1D}}{\partial x} \right|_{x^*}
		- b \, \tau_{\rm P}(\chi) 
		\left[ \cos{(\chi)} + \left. \frac{\partial \bar{y}}{\partial x} \right|_{x^*} \sin{(\chi)} \right] & = 0,
		\label{eq:Peierls1D_deriv} \\
	\left. \frac{\partial^2 V_{\rm P}^{\rm 1D}}{\partial x^2} \right|_{x^*} 
		- b \, \tau_{\rm P}(\chi) \left. \frac{\partial^2 \bar{y}}{\partial x^2} \right|_{x^*} \sin{(\chi)}  & = 0 .
		\label{eq:Peierls1D_deriv2}
\end{align}
This pair of equations defines the unstable position $x^*$ and the yield stress $\tau_{\rm P}(\chi)$. 
When the MRSSP coincides with the \hkl(-101) glide plane, \ie{} for $\chi=0$,
Eq. \ref{eq:Peierls1D_deriv2} shows that the unstable position is the inflexion point 
of the Peierls potential.  
The Peierls stress is the corresponding slope, \ie{} the maximum derivative of the Peierls potential, divided by $b$.
For any other MRSSP, the unstable position depends in principle also on the curvature of the dislocation 
trajectory.  However, \abinitio{} calculations have shown that this trajectory is smooth and not far from a straight line in its first part,
where the unstable position $x^*$ is located \cite{Dezerald2016}.
As illustrated in Fig. \ref{fig:trajectory}, the first part of the trajectory can thus be approximated by a straight line as $\bar{y}(x)=x\tan{(\alpha)}$,
where $\alpha$ is the angle between the approximated straight trajectory
and the \hkl(-101) glide plane\footnote{We define $\alpha$ here as the orientation of the tangent to the trajectory at the initial position, rather than the orientation of the straight line linking the initial position 
to the saddle point as in Refs. \cite{Dezerald2016,Kraych2019}, because a better agreement is obtained with this definition between the modified Schmid law (Eq. \ref{eq:PeierlsStress}) and the numerical solution of Eqs. \ref{eq:Peierls1D_deriv} and \ref{eq:Peierls1D_deriv2}.}.
With this approximation, $\partial^2 \bar{y}/\partial x^2=0$ and from Eq. \ref{eq:Peierls1D_deriv2}, $x^*$ remains at the Peierls potential 
inflexion point and does not depend on $\chi$.
Eq. \ref{eq:Peierls1D_deriv} then leads to the yield stress
\begin{align}
	\tau_{\rm P}(\chi) &= \frac{1}{b \left[ \cos{(\chi)} + \tan{(\alpha)} \sin{(\chi)} \right]}
		\left. \frac{\partial V_{\rm P}^{\rm 1D}}{\partial x} \right|_{x^*} 
	\nonumber \\
	&= \frac{\cos{(\alpha)}}{\cos{(\chi-\alpha)}}\tau_{\rm P}^0 ,
	\label{eq:PeierlsStress}
\end{align}
where the Peierls stress for $\chi=0$ is given by
\begin{equation*}
	\tau_{\rm P}^0 = \frac{1}{b}
		\left. \frac{\partial V_{\rm P}^{\rm 1D}}{\partial x} \right|_{x^*} .
\end{equation*}

\begin{figure}[!b]
	\begin{center}
		\includegraphics[width=0.7\linewidth]{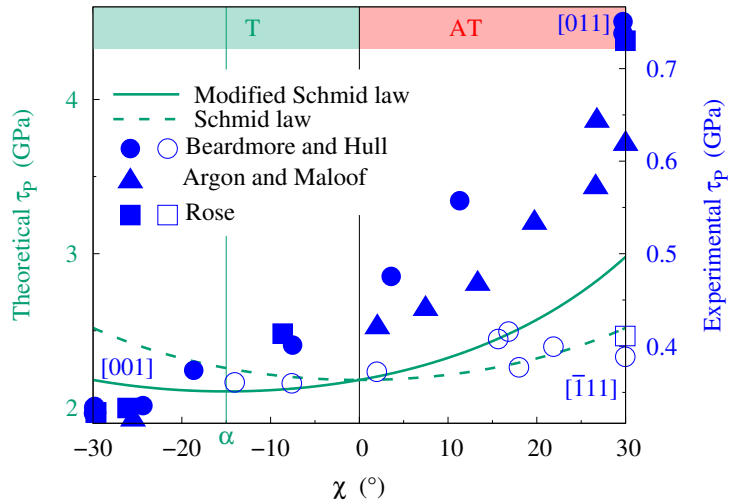}
	\end{center}
	\caption{Resolved shear stress for plastic yield in BCC tungsten
	as a function of the angle $\chi$ between the MRSSP and the \hkl{110} 
	glide plane.
	Theoretical values correspond to 0\,K \abinitio{} calculations using either Schmid law (dashed green line) or  
	including the T/AT asymmetry (solid green line) based on Eq. \ref{eq:PeierlsStress} with $\alpha=-15^{\circ}$.
	Experimental values \cite{Rose1962,Beardmore1965,Argon1966} were measured at 77\,K
	for a tensile axis between \hkl[001] and \hkl[011] (filled symbols)
	and between \hkl[001] and \hkl[-111] (open symbols).
	A scaling factor of 6 is used between theoretical (left coordinate axis)
	and experimental values (right axis).}
	\label{fig:tauP_chi}
\end{figure}

When the dislocation trajectory is straight and coincides with the macroscopic \hkl(-101) glide plane,
\ie{} when $\alpha=0$, one recovers the classical $1/\cos(\chi)$ Schmid law. However, when the trajectory is deviated as in Fig. \ref{fig:trajectory}, $\alpha\neq0$ and Eq. \ref{eq:PeierlsStress} shows that the shear stress 
which needs to be considered in the yield criterion is not the shear stress 
resolved in the \hkl(-101) glide plane, $\tau\cos{(\chi)}$, 
but rather the shear stress resolved in the plane tangent to the dislocation trajectory, $\tau \cos{(\chi-\alpha)}/\cos{(\alpha)}$.
The lowest Peierls stress is thus obtained when the MRSSP coincides with the tangent plane (\ie{} when $\chi=\alpha$), which  lies in the twinning region in all BCC transition metals as discussed above. We thus recover the T/AT asymmetry. Moreover, we find that the amplitude of the asymmetry increases with the norm of $\alpha$, and thus increases with the deviation of the dislocation trajectory from a straight line.

We compare in Fig. \ref{fig:tauP_chi} the yield stress predicted in tungsten 
from the modified Schmid law (Eq. \ref{eq:PeierlsStress}), using the angle $\alpha=-15^{\circ}$ measured on the dislocation trajectory
(Fig. \ref{fig:trajectory}), with experimental values obtained on single crystals at 77\,K for different orientations of the tensile axis 
\cite{Rose1962,Beardmore1965,Argon1966}
(see appendix \ref{app:chi_exp} for a derivation of the corresponding $\chi$ angles).
Because of the discrepancy between  theoretical and experimental 
Peierls stresses, and also because of the different temperatures 
(0\,K for \abinitio{} and 77\,K for experiments), theoretical and experimental data are shown with different scales. 
As already noted, the modified Schmid law correctly predicts that the twinning region with $\chi<0$ is easier to shear than the antitwinning region, in qualitative agreement with the experimental data. 
But experiments also show that the yield stress does not depend only on $\chi$:
the T/AT asymmetry is very strong for a tensile axis
between \hkl[001] and \hkl[011] (filled  symbols in Fig. \ref{fig:tauP_chi})
whereas the yield stress is almost constant 
between \hkl[001] and \hkl[-111] (open symbols in Fig. \ref{fig:tauP_chi}). 
The modified Schmid law given by Eq. \ref{eq:PeierlsStress} cannot account for such variations
between different orientations corresponding to the same MRSSP.  
We will show in the following that these variations partly arise from the second source of deviation from Schmid's law, non-glide stresses.

\subsection{Relaxation volume and tension-compression asymmetry}
\label{sec:relaxation_volume}

\begin{figure}[!b]
	\begin{center}
		\includegraphics[width=0.4\linewidth]{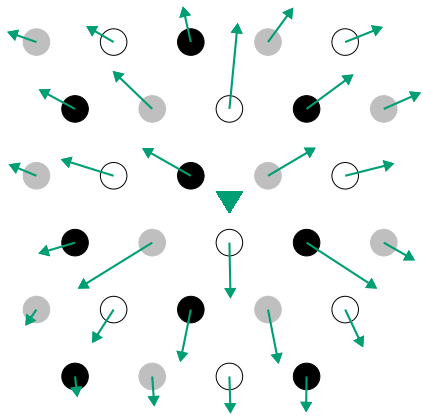}
	\end{center}
	\caption{Atomic displacements in the \hkl(111) plane perpendicular 
	to the dislocation line induced in BCC tungsten by the core dilatation 
	of a screw dislocation in its easy core configuration. 
	Displacement vectors
	have been magnified by a factor 50. 
	The dislocation position is indicated by a downward triangle.
	 }
	\label{fig:Uedge}
\end{figure}

\Abinitio{} calculations have shown that screw dislocations in BCC transition metals 
induce a short-range dilatation elastic field in addition to the elastic field 
given by the Volterra solution \cite{Clouet2009}.
This can be seen on the atomic displacements in Fig. \ref{fig:Uedge}, which have a component perpendicular to the dislocation line,
\ie{} an edge component, in the vicinity of the core. 
Part of this edge component could be a consequence of elastic anisotropy on the Volterra displacements, but tungsten is close to elastically isotropic. The edge displacements visible in Fig. \ref{fig:Uedge} are thus due to a 2D expansion centered at the dislocation core.
The elastic field induced by this dilatation is short ranged compared to the Volterra elastic field,
with a displacement varying as $1/r$ with $r$ the distance to the dislocation line instead of $\ln{(r)}$.
This dilatation partly arises from anharmonicity in the crystal response 
and also from the atomic structure of the dislocation core.
The core dilatation is responsible for the dislocation formation volume,
which manifests itself experimentally through an increase of the average lattice parameter
with the dislocation density \cite{Crussard1949}.

This core field can be modeled within elasticity using either line-force dipoles 
\cite{Gehlen1972,Hirth1973,Clouet2009,Clouet2011a,Clouet2011b}
or a 2D Eshelby inclusion \cite{Eshelby1957,Eshelby1959,Kraych2019},
both models being equivalent \cite{Clouet2018}.
Using the latter picture, a cylindrical inclusion 
of surface $S_0$ and eigenstrain tensor $\barT{\varepsilon}^*$
is associated to the dislocation core.
For the elastic field far from the dislocation and its coupling with an applied stress, 
it is actually sufficient to consider the relaxation volume tensor
$\barT{\Omega} = S_0 \, \barT{\varepsilon}^*$ defined per unit length of dislocation.
When the dislocation is in its ground state, the 3-fold symmetry around the \hkl[111] axis
of the easy core imposes that $\barT{\Omega}$ is diagonal with only two independent components:
$\barT{\Omega}=\diag{(\Omega_{11},\Omega_{11},\Omega_{33})}$.
Values of this tensor can be directly obtained from \abinitio{} calculations because, due to the relaxation volume, the dislocation energy varies when a stress tensor $\barT{\sigma}$
is applied, with a corresponding interaction energy 
$E^{\rm inter} = -\Omega_{ij} \sigma_{ij}$.
In the case of \abinitio{} modeling of a dislocation dipole with periodic boundary conditions,
Eq. \ref{eq:dE_epsi}, which gives the energy variation of the supercell 
caused by a homogeneous applied strain $\barT{\varepsilon}$, needs to be generalized to
\begin{equation}
	\Delta E(\barT{\varepsilon}) = \frac{1}{2} S \, C_{ijkl}\varepsilon_{ij}\varepsilon_{kl}
	+ C_{ijkl}(b_i A_j - 2\,\Omega_{ij}) \varepsilon_{kl},
	\label{eq:dE_epsi_relax}
\end{equation}
leading to a stress in the supercell
\begin{equation}
	\sigma_{ij}(\barT{\varepsilon}) 
		= C_{ijkl} \left( \varepsilon_{kl} + \frac{b_k A_l - 2\,\Omega_{kl}}{S} \right).
	\label{eq:stress_epsi_relax}
\end{equation}
Thanks to this expression, the relaxation volume can be directly deduced from the stress in the supercell after energy relaxation. For tungsten, we obtained  $\Omega_{11}=9$ and $\Omega_{33}=-4$\,\AA$^3$ 
per Burgers vector of screw dislocation. 
A similar dilatation perpendicular to the dislocation line and contraction along the line was obtained in other BCC metals \cite{Dezerald2014}.

\begin{figure}[!bth]
	\begin{center}
		\includegraphics[width=0.8\linewidth]{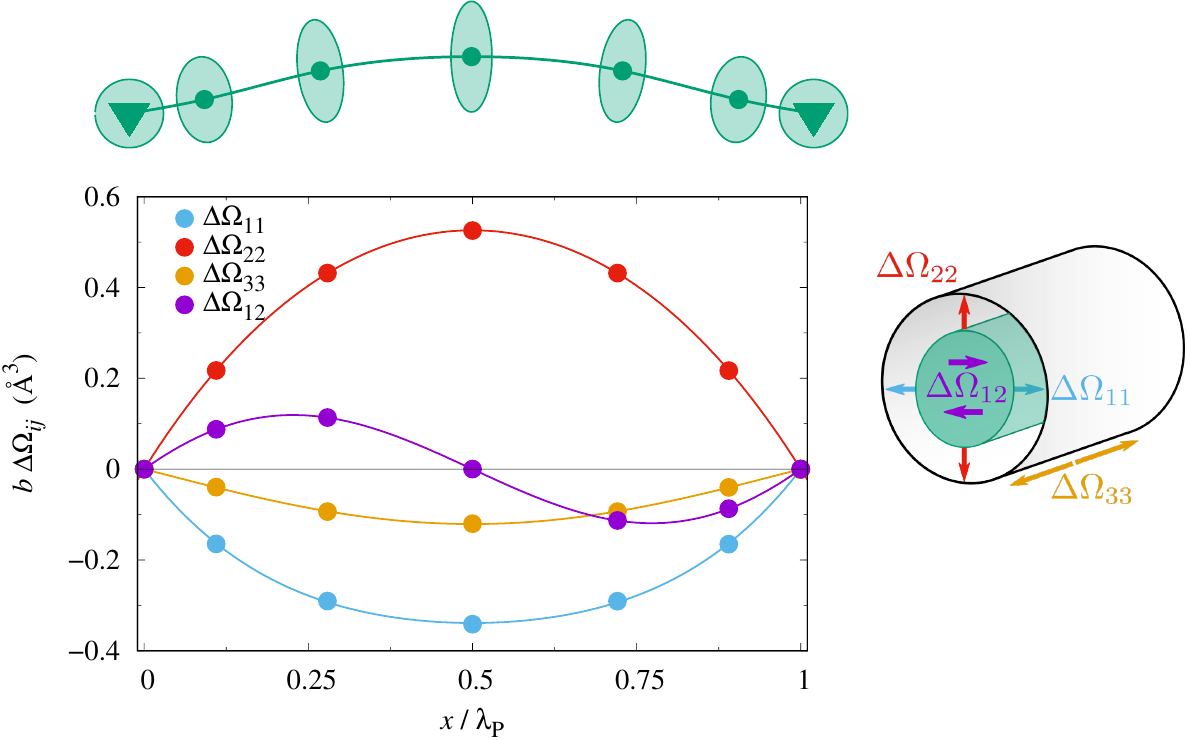}
	\end{center}
	\caption{Variation of the relaxation volume tensor along the minimum energy path between easy core configurations.
	The different tensor components are shown as a function of the dislocation position $x$ along the glide direction 
	normalized by the distance $\lambda_{\rm P}$ between Peierls valleys. 
	The corresponding shape of the 2D Eshelby inclusion along this path
	is sketched in the upper part.
	}
	\label{fig:coreVol}
\end{figure}

The variation of the dislocation relaxation volume along the minimum energy path between easy cores, $\Delta\barT{\Omega}$, is actually more important for plasticity than the absolute value of the tensor.
Using the stress variation measured during \abinitio{} NEB calculations 
of the Peierls energy barrier (\cf{} section \ref{sec:PeierlsStress}), one obtains the variation of the relaxation volume tensor
shown in Fig. \ref{fig:coreVol}.
Except in the initial and final positions, the 3-fold symmetry is not obeyed 
along the path and the relaxation volume tensor has now the following form
in the Cartesian basis associated with the dislocation (see Fig. \ref{fig:core} and the sketch on the right hand-side of Fig. \ref{fig:coreVol}):
\begin{equation*}
	\barT{\Omega} = 
	\begin{pmatrix}
		\Omega_{11} & \Omega_{12} & 0 \\
		\Omega_{12} & \Omega_{22} & 0 \\
		0           & 0           & \Omega_{33}
	\end{pmatrix}.
\end{equation*}
There is no symmetry argument imposing that the components $\Omega_{13}$ and $\Omega_{23}$
should be null, but \abinitio{} calculations show that this is the case for tungsten \cite{Kraych2019} and other BCC transition metals.
As described in appendix \ref{app:elastic}, a slight adjustment of the elastic constants 
is necessary to impose that the initial and final positions along the path, both corresponding
to the same ground state, have the same relaxation volume. 
This correction is reasonable considering the precision of the elastic constants from \abinitio{} calculations
and is also compatible with the variations of the elastic constants in the dislocated crystal. 

Because of the variations of the relaxation volume and its coupling with the applied stress tensor, the enthalpy of the dislocation
as a function of its position in the \hkl(111) plane in Eq. \ref{eq:Peierls2D} becomes
\begin{equation}
	\Delta H^{\rm 2D}_{\rm P}(x,y) = V_{\rm P}^{\rm 2D}(x,y)
		- \sigma_{ij} \, \Delta\Omega_{ij}^{\rm 2D}(x,y)
		- \tau_{yz}\,b\,x + \tau_{xz}\,b\,y.
	\label{eq:Peierls2D_relax}
\end{equation}
Since the dislocation trajectory is not sensitive to the applied stress tensor,
(\cf{} section \ref{sec:trajectory}), one can recover a 1D functional 
for the enthalpy barrier between Peierls valleys
\begin{equation}
	\Delta H^{\rm 1D}_{\rm P}(x) = V_{\rm P}^{\rm 1D}(x)
		- \sigma_{ij} \, \Delta\Omega_{ij}^{\rm 1D}(x)
		 - \tau\,b \left[ x \cos{(\chi)} + \bar{y}(x) \sin{(\chi)} \right],
	\label{eq:Peierls1D_relax}
\end{equation}
where $\Delta\barT{\Omega}^{\rm 1D}(x) = \Delta\barT{\Omega}^{\rm 2D}(x,\bar{y}(x))$ is the variation of the relaxation volume 
along the minimum energy path.
Finally, keeping our approximation of a straight dislocation trajectory defined by an angle $\alpha$, one obtains the yield stress
\begin{equation}
	\tau_{\rm P}(\chi,\barT{\sigma})
		= \frac{1}{b \left[ \cos{(\chi)} + \tan{(\alpha)} \sin{(\chi)} \right]}
		\left(
		\left. \frac{\partial V_{\rm P}^{\rm 1D}}{\partial x} \right|_{x^*(\barT{\sigma})} 
		-
		\left. \sigma_{ij}\frac{\partial \Delta\Omega_{ij}^{\rm 1D}}{\partial x} \right|_{x^*(\barT{\sigma})} 
		\right)
	\label{eq:PeierlsStress_relax}
\end{equation}
where $x^*(\barT{\sigma})$ is the inflexion point on the generalized Peierls potential
$V_{\rm P}^{\rm 1D}(x) - \sigma_{ij} \, \Delta\Omega_{ij}^{\rm 1D}(x)$, which depends \textit{a priori} on the non-glide stress $\barT{\sigma}$. Using a first-order expansion in $\barT{\sigma}$ of the two instability conditions defining the Peierls stress, 
$\partial \Delta H_{\rm P}^{\rm 1D}/\partial x=0$ and
$\partial^2 \Delta H_{\rm P}^{\rm 1D}/\partial x^2=0$,
one shows \cite{Kraych2019} that the variation of the inflexion point position and its impact on the derivatives 
appearing in Eq. \ref{eq:PeierlsStress_relax} can be neglected finally leading to
\begin{equation}
	\tau_{\rm P}(\chi,\barT{\sigma})
	= \frac{\cos{(\alpha)}}{\cos{(\chi-\alpha)}}
	\left ( \tau_{\rm P}^0 - \sigma_{ij} \, \Delta\Omega_{ij}' \right),
\end{equation}
with $\Delta\barT{\Omega}'$ the derivative of the relaxation volume tensor
calculated at the inflexion point of the Peierls potential $V_{\rm P}^{\rm 1D}(x)$.

\begin{figure}[!bth]
	\begin{center}
		\includegraphics[width=0.49\linewidth]{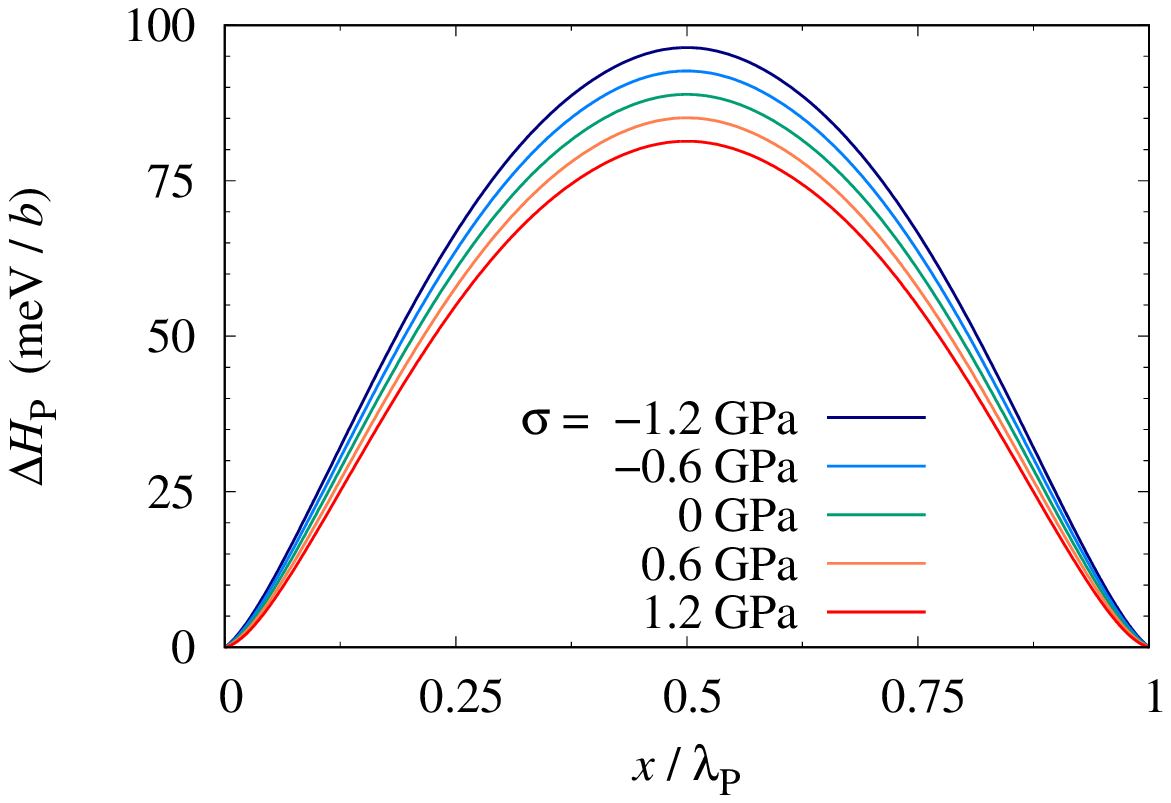}
		\hfill
		\includegraphics[width=0.49\linewidth]{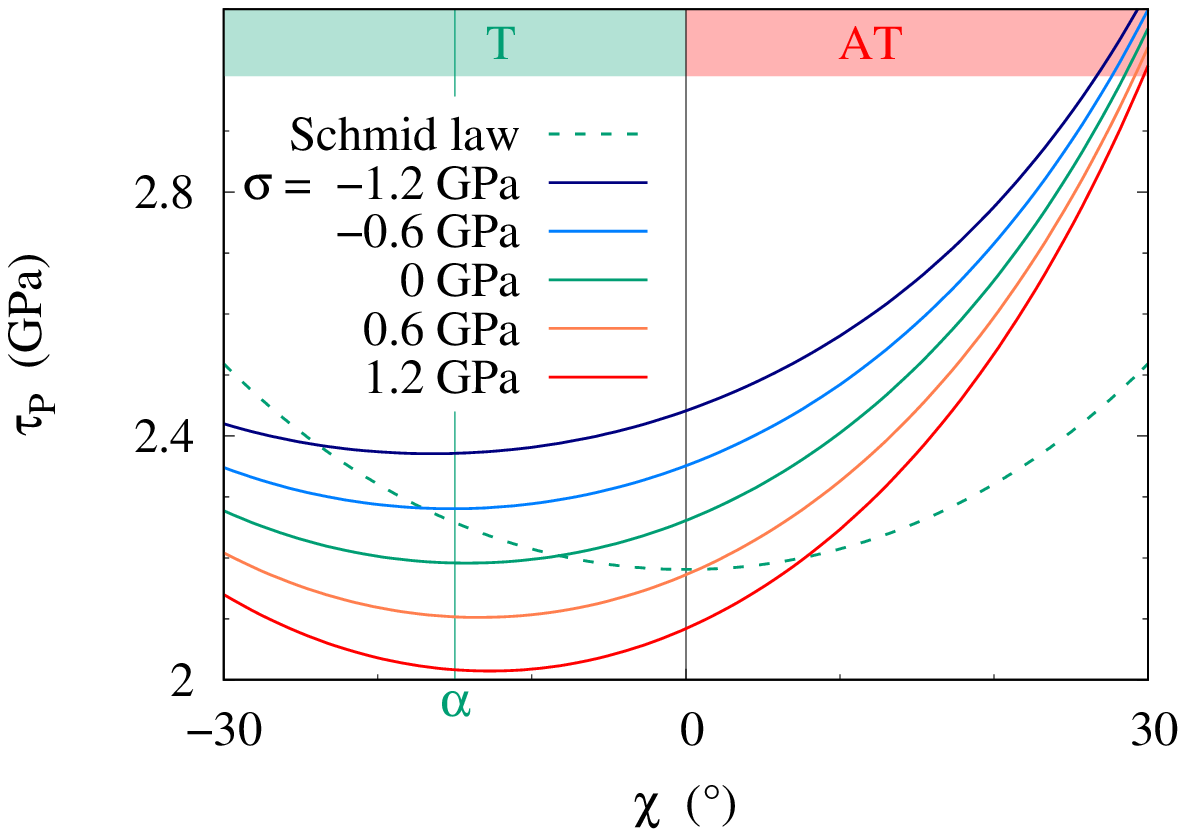}
	\end{center}
	\caption{Dependence of the Peierls enthalpy barrier (left) and yield stress (right) on a non-glide stress $\sigma$ perpendicular to the MRSSP 
	(\cf{} Eq. \ref{eq:tauSigmaStess} for the stress tensor definition). $\chi$ is the angle between the MRSSP and the $\hkl{110}$ glide plane
	and is null in the left figure.  }
	\label{fig:Peierls_relax}
\end{figure}

To illustrate the effect of the relaxation volume on the Peierls stress, we consider a mechanical loading composed of a shear stress $\tau$ 
in a MRSSP defined by its angle $\chi$, as in previous section,
combined with a traction/compression stress $\sigma$ perpendicular to the dislocation line,
leading to the following stress tensor in the dislocation axis
\begin{equation}
	\barT{\sigma} = 
	\begin{pmatrix}
		-\sigma \cos{(2\chi)} & -\sigma \sin{(2\chi)} &  \tau \sin{(\chi)} \\
		-\sigma \sin{(2\chi)} &  \sigma \cos{(2\chi)} & -\tau \cos{(\chi)} \\
		\tau \sin{(\chi)}     & -\tau \cos{(\chi)}    & 0
	\end{pmatrix}.
	\label{eq:tauSigmaStess}
\end{equation}
For this loading, the enthalpy variation and associated yield stress are given by
\begin{multline}
	\Delta H^{\rm 1D}_{\rm P}(x) = V_{\rm P}^{\rm 1D}(x)
		+ \sigma \cos{(2\chi)} \left[ \Delta\Omega_{11}(x) - \Delta\Omega_{22}(x) \right]
		+ 2 \sigma \sin{(2\chi)} \Delta \Omega_{12}(x)
		\\
		 - \tau\,b \left[ x \cos{(\chi)} + \bar{y}(x) \sin{(\chi)} \right],
	\label{eq:Peierls1D_relax_traction}
\end{multline}
and
\begin{equation}
	\tau_{\rm P}(\chi,\sigma)
	= \frac{\cos{(\alpha)}}{\cos{(\chi-\alpha)}}
	\left ( \tau_{\rm P}^0  
		+ \sigma \cos{(2\chi)} \left[ \Delta\Omega_{11}' - \Delta\Omega_{22}' \right]
		+ 2 \sigma \sin{(2\chi)} \Delta \Omega_{12}'
	\right).
\end{equation}
The yield stress is now sensitive to the anisotropy of the dislocation core dilatation, 
or more precisely to its variation along the dislocation migration path.
When the MRSSP is in tension ($\sigma>0$), both the enthalpy barrier
and the yield stress decrease (Fig. \ref{fig:Peierls_relax}) 
as the dislocation core tends to expand in the direction perpendicular to the glide plane ($\Delta\Omega_{22}>0$)
and to contract in the glide direction ($\Delta\Omega_{11}<0$) when transitioning between  Peierls valleys (see Fig. \ref{fig:coreVol}).
The variation of the dislocation relaxation volume therefore explains 
the observed decrease of the Peierls stress when the MRSSP is in tension,
a general feature of the departure from Schmid's law observed in BCC metals modeled with various atomistic models \cite{Vitek2004,Groger2008,Chen2013,Groger2014,Hale2015}. 

\subsection{Generalized yield criterion}
\label{sec:generalized_yield}

\begin{figure}[bth]
	\begin{center}
		\includegraphics[width=0.70\linewidth]{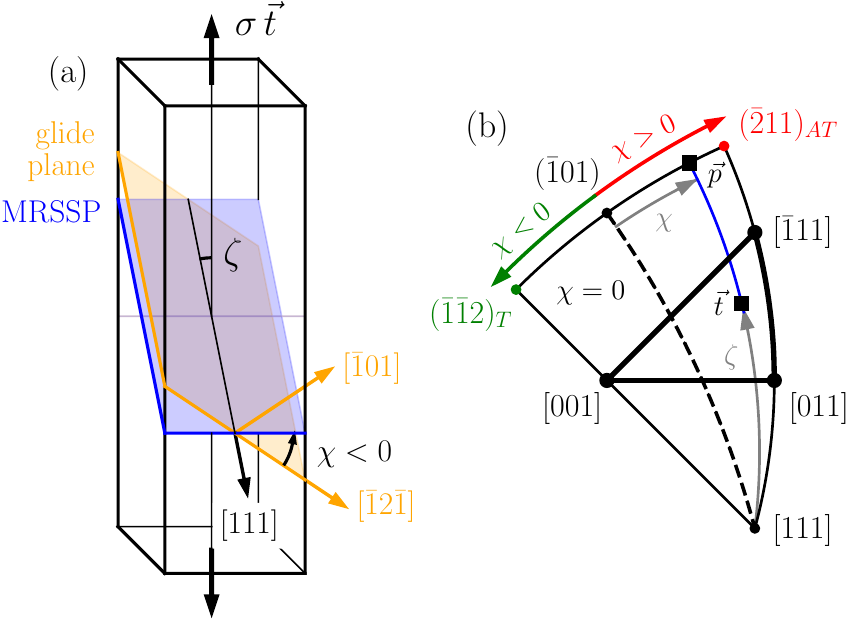}
	\end{center}
	\caption{(a) Sketch of a tensile mechanical test showing the \hkl(-101) glide plane and the maximum resolved shear stress plane (MRSSP). (b) Angles $\zeta$ and $\chi$ defining the orientation of the tensile axis $\vec{t}$ in the standard stereographic projection of the minimum irreducible zone of the \hkl[111]\hkl(-101) slip system with $\zeta \in{[0^\circ;\,90^\circ]}$ and $\chi \in{[-30^\circ;\,+30^\circ]}$. The thick black triangle delimited by \hkl[001], \hkl[011] and \hkl[-111] is the standard stereographic triangle where the \hkl[111]\hkl(-101) slip system has the highest Schmid factor. 
	}
	\label{fig:SchemaAngles}
\end{figure}

We study now how the coupling of the applied stress with 
both the trajectory and the relaxation volume of the $1/2\hkl<111>$ screw dislocation 
impacts the yield stress for the simplest mechanical loading:
a traction or compression test on a single crystal.
Under a uniaxial loading of magnitude $\sigma$ along an axis $\vec{t}$, the stress tensor applied to the crystal is:
\begin{equation}
\barT{\Sigma}=\sigma (\vec{t}{\otimes}\vec{t}). \label{eq:uniaxial}
\end{equation}
Using spherical coordinates to project the tensile axis $\vec{t}$ in the frame of the gliding dislocation, the stress tensor $\overline{\overline{\Sigma}}$ is expressed as:
\begin{equation}
	\barT{\Sigma} = \sigma
	\begin{bmatrix} 
	\sin^2{(\zeta)}\sin^2{(\chi)} & \sin^2{(\zeta)}\sin{(2\chi)}/2 & \sin{(2\zeta)}\sin{(\chi)}/2 \\
 & \sin^2{(\zeta)}\cos^2{(\chi)} & \sin{(2\zeta)}\cos{(\chi)}/2 \\
 &  & \cos^2{(\zeta)}
\end{bmatrix},
\label{eq:stress_tensor}
\end{equation}
where $\zeta$ is the angle between the slip direction, \ie{} the Burgers vector $\vec{b}$, and the tensile axis $\vec{t}$, 
and $\chi$ the angle between the glide plane and the MRSSP\footnote{Following  Duesbery \cite{Duesbery1984}, 
$\chi$ can also be defined as the angle between the normal $\vec{n}$ to the glide plane 
and the projection $\vec{p}$ of the tensile axis on the plane orthogonal to the slip direction.}
(Fig. \ref{fig:SchemaAngles}). The enthalpy is now expressed as
\begin{equation}
\begin{aligned}
\Delta H_{\rm P}^{\rm 1D}(x)={} 
	& V_{\rm P}^{\rm 1D}(x) 
	-\frac{1}{2} \sigma \sin{(2\zeta)} \, b \left[ x\cos{(\chi)} + \bar{y}(x)\sin{(\chi)} \right] \\
	& - \frac{1}{2} \sigma \sin^2{(\zeta)} \left\{ 
	[\Delta \Omega_{22}(x) - \Delta \Omega_{11}(x)] \cos{(2\chi)}
		+ 2 \Delta \Omega_{12}(x)\sin{(2\chi)} \right\} \\
	& - \frac{1}{2} \sigma \sin^2{(\zeta)} [\Delta \Omega_{11}(x) + \Delta \Omega_{22}(x) + \Delta \Omega_{33}(x)]
	+ \frac{1}{2} \sigma [1 - 3 \cos^2{(\zeta)}] \Delta \Omega_{33}(x) .
\label{eq:non_glide}
\end{aligned}
\end{equation}
The tensile yield stress at 0\,K, $\sigma^0_Y$, is again found at a position $x^*$ satisfying the instability conditions $\partial \Delta H_{\rm P}^{\rm 1D}/\partial x=0$ and
$\partial^2 \Delta H_{\rm P}^{\rm 1D}/\partial x^2=0$, which need to be solved numerically for each orientation of the loading
axis defined by $\zeta$ and $\chi$, thus
leading to $\sigma^0_{\rm Y}(\zeta,\chi)$.

\begin{figure}[bth]
	\begin{center}
		\includegraphics[width=0.8\linewidth]{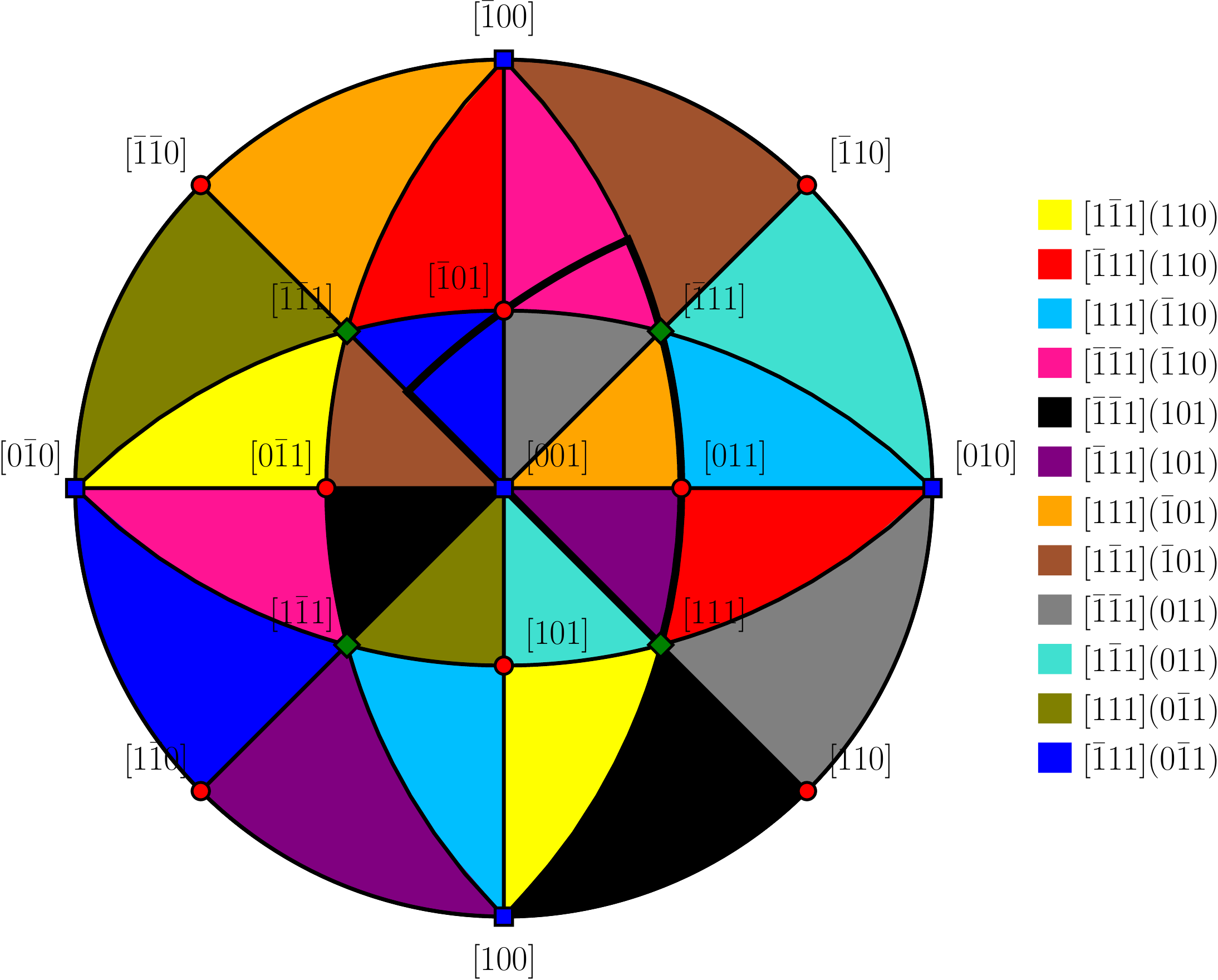}
	\end{center}
	\caption{Stereographic projection of the BCC lattice showing in colors the regions where each individual \hkl<111>\hkl{110} slip system has the highest Schmid factor. 
	The thick black triangle is the  minimum  irreducible  zone  of the \hkl[111]\hkl(-101) slip system with $\zeta \in \left[ 0\,,90^\circ \right]$ 
	and $\chi \in \left[ -30^\circ\,,30^\circ \right]$.
	}
	\label{fig:StereographicCircleSchmid}
\end{figure}

With non-Schmid effects taken into account, it becomes necessary to consider all possible \hkl<111>\hkl{110} slip systems to evaluate the minimum tensile yield stress as a function of the orientation of the loading axis. The twelve \hkl<111>\hkl{110} slip systems of the BCC lattice,
defined by 4 different \hkl<111> slip directions and 3 different \hkl{110} glide planes,
are presented in the stereographic projection in Fig. \ref{fig:StereographicCircleSchmid}.
In any region of the stereographic projection delimited by \hkl<100>, \hkl<110>, and \hkl<111> orientations, a single slip system has a maximum Schmid factor. 
This defines the standard stereographic triangle for this slip system.
Among the equivalent slip systems at each corner of a stereographic triangle, 
all  make the same angle $\zeta$ with the tensile axis, but half of them have a positive $\chi$ angle, and the other half a negative $\chi$ angle. Hence only half are sheared in the twinning sense, and the other half in the anti-twinning sense. 
With the $1/\cos{(\chi-\alpha)}$ variation of the yield stress predicted by the modified Schmid law, slip systems with positive or negative $\chi$ angles are no longer equivalent 
and systems with a tensile axis oriented towards negative $\chi$ (respectively positive $\chi)$
are easier to activate in tension (respectively in compression). 
Hence, there is a splitting of the slip systems into twinned and anti-twinned groups.
As a consequence, a full description of yield stress variation 
with tensile axis orientation cannot be restricted
to a single stereographic triangle, but two adjacent triangles are needed.

\begin{figure}[bth]
	\begin{center}
        \includegraphics[trim = 36mm 28mm 2mm 24mm, clip, width=1\linewidth]{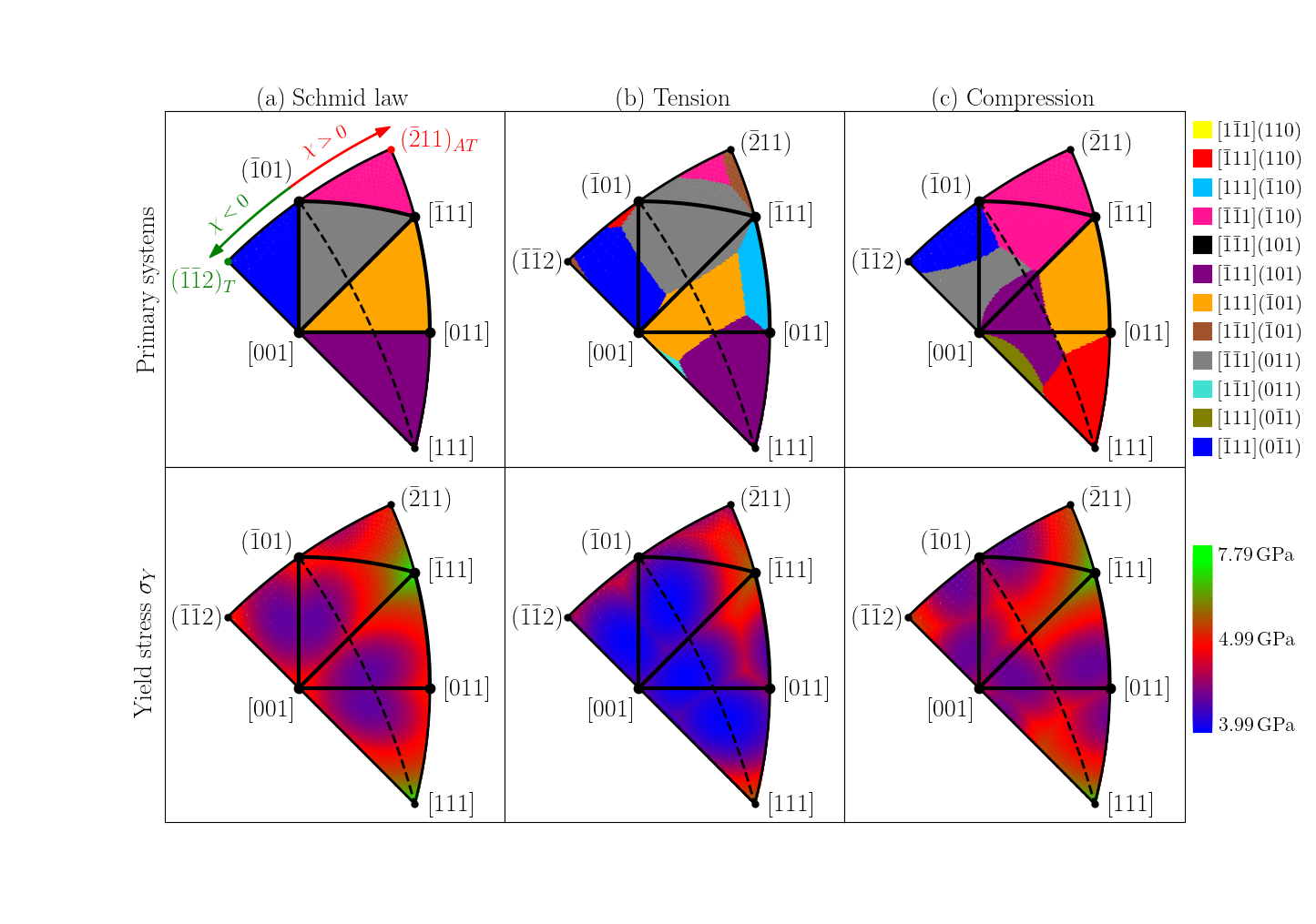}
	\end{center}
	\caption{Primary \hkl<111>\hkl{110} slip system (first row) and corresponding yield stress (second row) activated in tungsten single crystals under uniaxial loading at 0\,K as a function of the orientation of the deformation axis and predicted using the Schmid law (a), and the yield criterion developed in this work in tension (b) and compression (c). The color coding of the different slip systems and the stress range are indicated in the legend on the left.}
	\label{fig:PrimarySystemsLevel}
\end{figure}

The yield stress of each slip system was evaluated numerically at 0\,K as a function of the orientation of the loading axis $\vec{t}$. 
We present in the first row of Fig. \ref{fig:PrimarySystemsLevel} a color map showing the distribution of primary \hkl<111>\hkl{110} slip systems, {\ie} the systems having the lowest yield stress, as a function of the tensile axis orientation. The results are shown according to Schmid's law, and using the yield criterion including non-Schmid effects in tension and compression. The corresponding minimum yield stress is presented in the second row of Fig.\ref{fig:PrimarySystemsLevel}.
With Schmid's law, only one primary slip system exists in each stereographic triangle 
and the yield stress follows directly the distribution of Schmid factors, 
with easier glide in the regions near the $\hkl[001]-\hkl[011]$ edge and the $\chi=0$ line, 
and a maximum yield stress near the \hkl[-111] corner.
When non-Schmid effects are taken into account, several primary slip systems appear inside the $\hkl[001]-\hkl[011]-\hkl[-111]$ triangle, with different distributions in tension and compression. The region of the stereographic projection where the \hkl[111]\hkl(-101) system is activated
(orange region in Fig.\ref{fig:PrimarySystemsLevel}) 
is shifted towards $\chi<0$ in tension, and towards $\chi>0$ in compression. 
This shift of the primary slip system is also responsible for the emergence
of neighboring primary systems close to the edges of the stereographic triangle.
Looking now to the yield stresses (lower row of Fig. \ref{fig:PrimarySystemsLevel}), a tension / compression asymmetry appears clearly.
With a larger blue region corresponding to the lowest values, the average yield stress necessary to activate plasticity is lower in tension than in compression. 
This asymmetry is a direct consequence of the dislocation relaxation volume
and is mainly driven by the sign of the difference $\Delta\Omega_{22}-\Delta\Omega_{11}$
as analyzed in the previous section.
The generalized yield criterion which considers the coupling of the dislocation relaxation volume 
with the applied stress thus offers a physical explanation to the experimental observation
that the yield stress for a given angle $\zeta$ is generally higher in compression than in tension,
regardless of the orientation of the loading axis \cite{Byron1967,Liu1972,Takeuchi1972,Nawaz1975}.

\begin{figure}[bth]
	\begin{center}
		\includegraphics[width=1\linewidth]{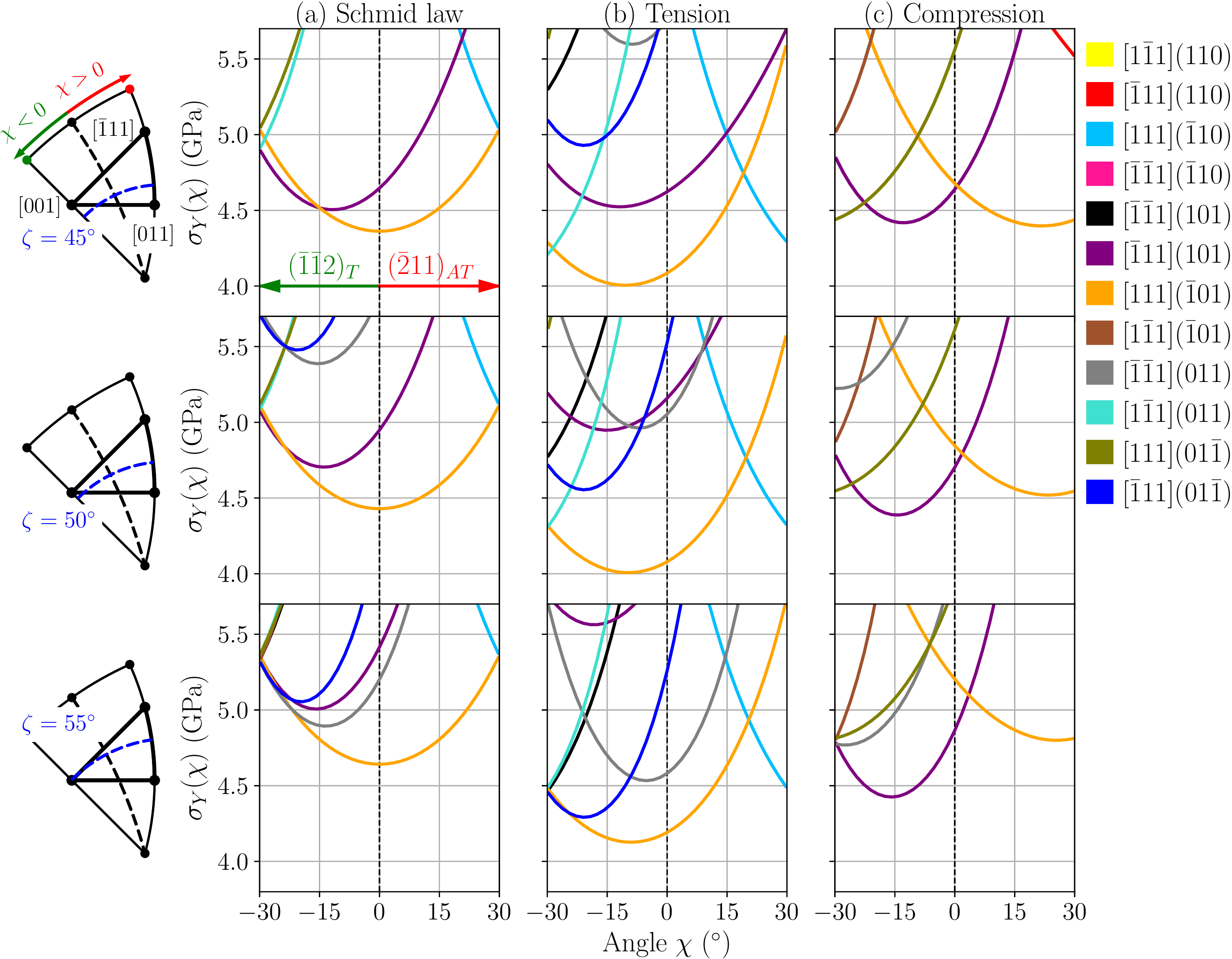}
	\end{center}
	\caption{Yield stress of the different slip systems	in tungsten under uniaxial loading at 0\,K 	as a function of the angle $\chi$ 	for $\zeta=45^\circ,\,50^\circ,\,55^\circ$:	(a) according to the Schmid law and including non-Schmid effects in (b) tension and (c) compression.
	}
	\label{fig:ThetaLines}
\end{figure}

To better visualize consequences of the modified Schmid law on the competition between different slip systems, profiles of the yield stress at 0\,K as a function of the orientation of the loading axis are presented in Fig. \ref{fig:ThetaLines} for three different constant $\zeta$ cuts of the stereographic projection. For each slip system, a clear T/AT asymmetry is visible. It also appears that the yield stress is lower in tension than in compression, whatever the orientation of the loading axis, and that the activated primary slip system usually varies between tension and compression for the same orientation.

\begin{figure}[bth]
	\begin{center}
		\includegraphics[width=1\linewidth]{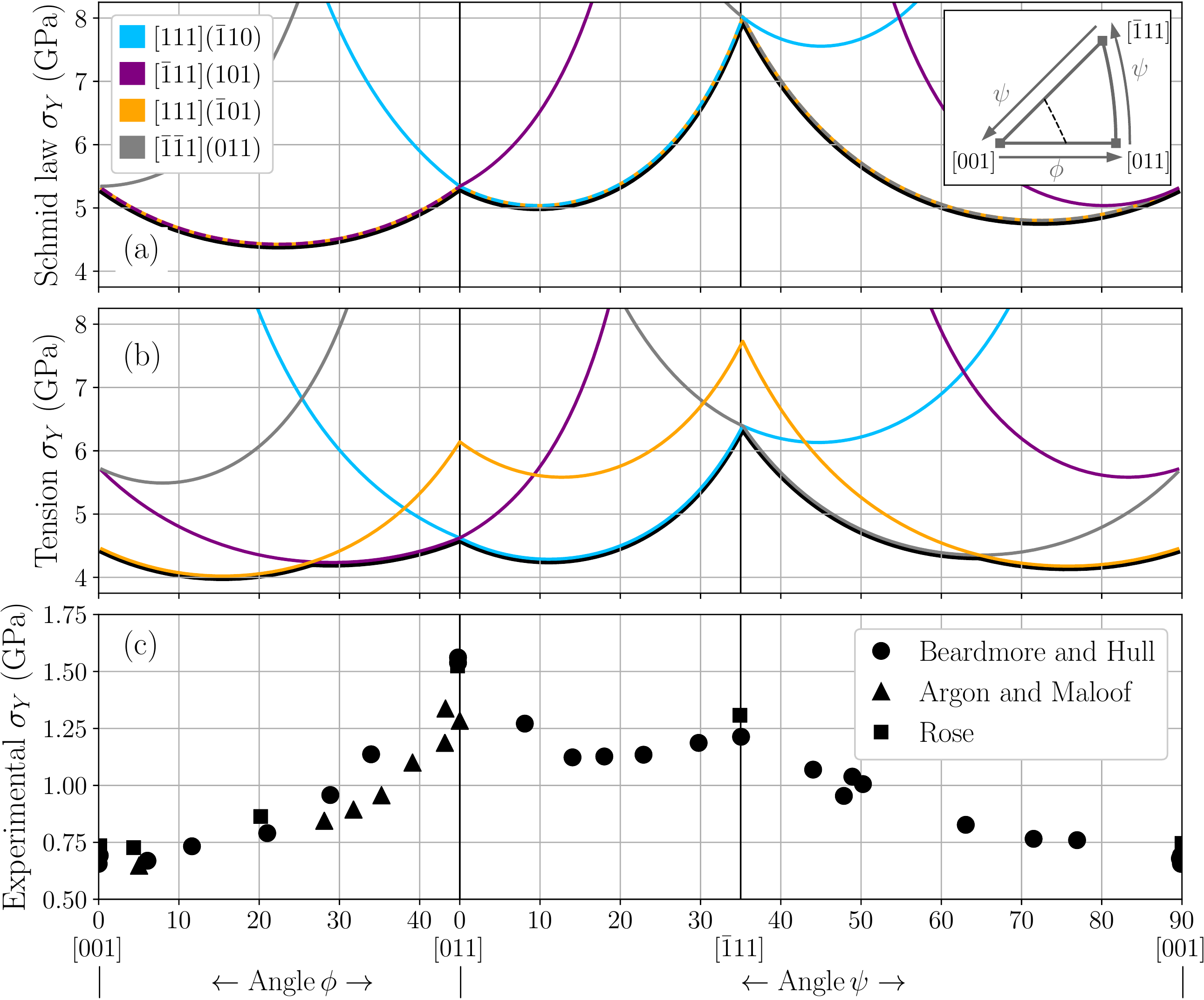}
	\end{center}
	\caption{Variation of the minimum yield stress (dashed black lines) among all \hkl<111>\hkl{110} slip systems at 0\,K according to (a) the Schmid law and (b) the yield criterion developed in this work, as a function of the orientation of the tensile axis along the $\hkl[001]-\hkl[011]-\hkl[-111]$ edges of the standard stereographic triangle. The path is sketched in the inset of (a) with the definition of the angles $\phi$ and $\psi$. (c) Experimental data at 77\,K in tension from Beardmore and Hull ($\bullet$) \cite{Beardmore1965}, Argon and Maloof ($\blacktriangle$) \cite{Argon1966}, and Rose \textit{et al}. ($\blacksquare$) \cite{Rose1962}.}
	\label{fig:EdgesTriangle_b111}
\end{figure}

We compare in Fig. \ref{fig:EdgesTriangle_b111} the predicted yield stress at 0\,K for tungsten with various experimental data \cite{Beardmore1965,Rose1962,Argon1966} obtained in tension at 77\,K for single crystals oriented along the edges of the standard stereographic triangle.
As it will be shown in the next section, the 77\,K temperature of experiments
is sufficiently low compared to tungsten melting temperature to have a minor impact 
on the yield stress orientation dependence.
Acknowledging the discrepancy between theoretical and experimental Peierls stresses 
discussed in section \ref{sec:PeierlsStress}, we only compare relative variations.
If the theoretical yield criterion correctly predicts a lower yield stress for \hkl[001]
than \hkl[-111] orientations, a feature actually already present in Schmid's law, 
it fails to predict the strong increase of the yield stress close to the \hkl[011] axis.
In particular along the $\hkl[001] - \hkl[011]$ edge, the theoretical criterion predicts an almost flat variation 
of the yield stress because of the competition between two primary slip systems, 
instead of a steep increase when approaching \hkl[011].
Although the modified Schmid law allows to rationalize the T/AT and tension/compression asymmetries,
it apparently still misses some ingredients to fully account for the yield stress variations
in all regions of the stereographic projection, in particular in the vicinity of the \hkl[011] direction. 
In this region, glide of $1/2\,\hkl<111>$ dislocations on \hkl{112} instead of \hkl{110} planes
has been evidenced experimentally, both from slip traces analysis on strained single crystals \cite{Argon1966}
and \insitu{} TEM straining experiments \cite{Caillard2018}.
Although no precise atomistic mechanism has been proposed until now to explain glide in \hkl{112} planes, 
the inclusion of \hkl<111>\hkl{112} slip systems in the yield criterion appears as a necessary
next step to fully describe the yield surface of BCC metals, 
in particular tungsten. 
As inclusion of additional slip systems could only lower the theoretical yield stress, 
one should probably also invoke a locking of the active $1/2\,\hkl<111>\hkl{110}$ slip systems 
to rationalize the steep increase of the yield stress experimentally observed close to the \hkl[011] direction.

\section{Thermal activation}

\begin{figure}[bth]
	\begin{center}
		\includegraphics[width=0.95\linewidth]{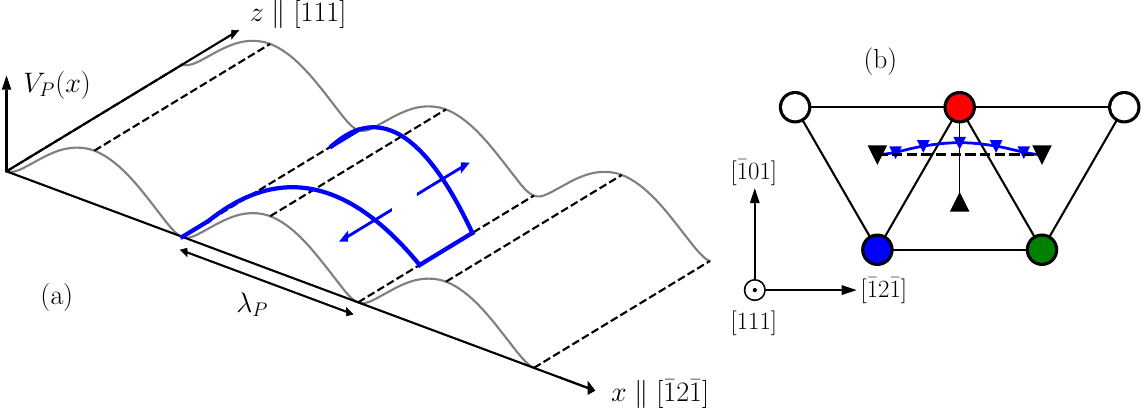}
	\end{center}
	\caption{(a) Sketch of a dislocation line (blue) crossing a Peierls barrier through the nucleation and propagation of a kink pair. 
	(b) Projection of the $1/2\,\hkl[111]$ screw dislocation trajectory gliding in the \hkl(-101) plane with the three most displaced \hkl[111] atomic columns  represented in blue, red and green.}
	\label{fig:KinkPropagation}
\end{figure}

We now describe how the effect of temperature on the yield stress 
can be modeled using \abinitio{} calculations.
At low temperature, glide of screw dislocations in BCC metals 
is thermally activated and operates through the nucleation of kink pairs across the Peierls barrier and their subsequent propagation along the dislocation line, as sketched in Fig. \ref{fig:KinkPropagation}(a).
In pure BCC metals, kinks glide along a dislocation line 
with a negligible lattice friction and the motion of the screw dislocation is controlled
by kink nucleation. Modeling kinked dislocations in pure BCC metals like tungsten
requires supercells too large for \abinitio{} calculations.
We therefore employ a multiscale approach based on a line tension model adjusted on \abinitio{} calculations
\cite{Proville2013}.

\subsection{Line tension model for kink-pair formation}

In the frame of the line tension (LT) approximation, the dislocation is modeled as an elastic line, 
whose ease to bend  is defined by its line tension. 
The shape of the dislocation is represented by a function $x(z)$ defining the position of the dislocation in its glide plane
as a function of the coordinate $z$ along the dislocation line (see Fig. \ref{fig:KinkPropagation}). 
The bow out of the dislocation line under an applied stress tensor $\barT{\Sigma}$ 
necessary to form a kink pair causes
a change in the dislocation enthalpy given by \cite{Proville2013}
\begin{equation}
H_{\rm LT}[x(z),\barT{\Sigma}]
= \int{ \ud{z} \left\{ \Delta H_{\rm P}^{\rm 1D}[x(z),\barT{\Sigma}] + \dfrac{\Gamma}{2} \left( \dfrac{\partial x}{\partial z} \right)^2 \right\} }, 
	\label{eq:LT}
\end{equation}
where $\Delta H_{\rm P}^{\rm 1D}$ is the enthalpy variation of the straight dislocation per unit length under the applied stress $\barT{\Sigma}$ (Eq. \ref{eq:Peierls1D_relax})
and $\Gamma$ the line tension, which is assumed isotropic and independent of the applied stress. 

The line tension $\Gamma$ can be extracted from \abinitio{} calculations following the approach of Proville \etal{} \cite{Proville2013,Dezerald2015}.
We consider a $2b$-wide supercell constructed by stacking two one-$b$ slabs, each containing a relaxed screw dislocation dipole in its easy core configuration. As the dislocation crosses the Peierls barrier in a \hkl{110} plane, the three \hkl<111> atomic columns defining the dislocation core and represented with colors in Fig. \ref{fig:KinkPropagation}(b) move parallel to the dislocation line along the \hkl[111] direction. To emulate the bow out of the line in the $2b$ simulation cell, the position of the dislocation in the lower $1b$-slab is kept fixed in its Peierls valley by freezing the displacement of the three core \hkl<111> columns, while a constrained displacement is imposed to these columns in the upper slab to mimic the beginning of kink nucleation. The change in energy resulting from the bow-out of the dislocation line under zero applied stress is then fitted to a discretized version of Eq. \ref{eq:LT} to extract the line tension $\Gamma$. 
Calculations in different BCC metals \cite{Dezerald2015} show only small metal-to-metal variations, in contrast with the line tension calculated from elasticity theory. 
The line tension defining the bow out of a kinked screw dislocation 
corresponds to a localized energy variation, which cannot be modeled with elasticity and for which an atomic description is needed. 
For BCC tungsten, we find $\Gamma = 3.41$\,eV/{\AA}.

\begin{figure}[bth]
	\begin{center}
	    \hspace*{-2mm}
		\includegraphics[width=1\linewidth]{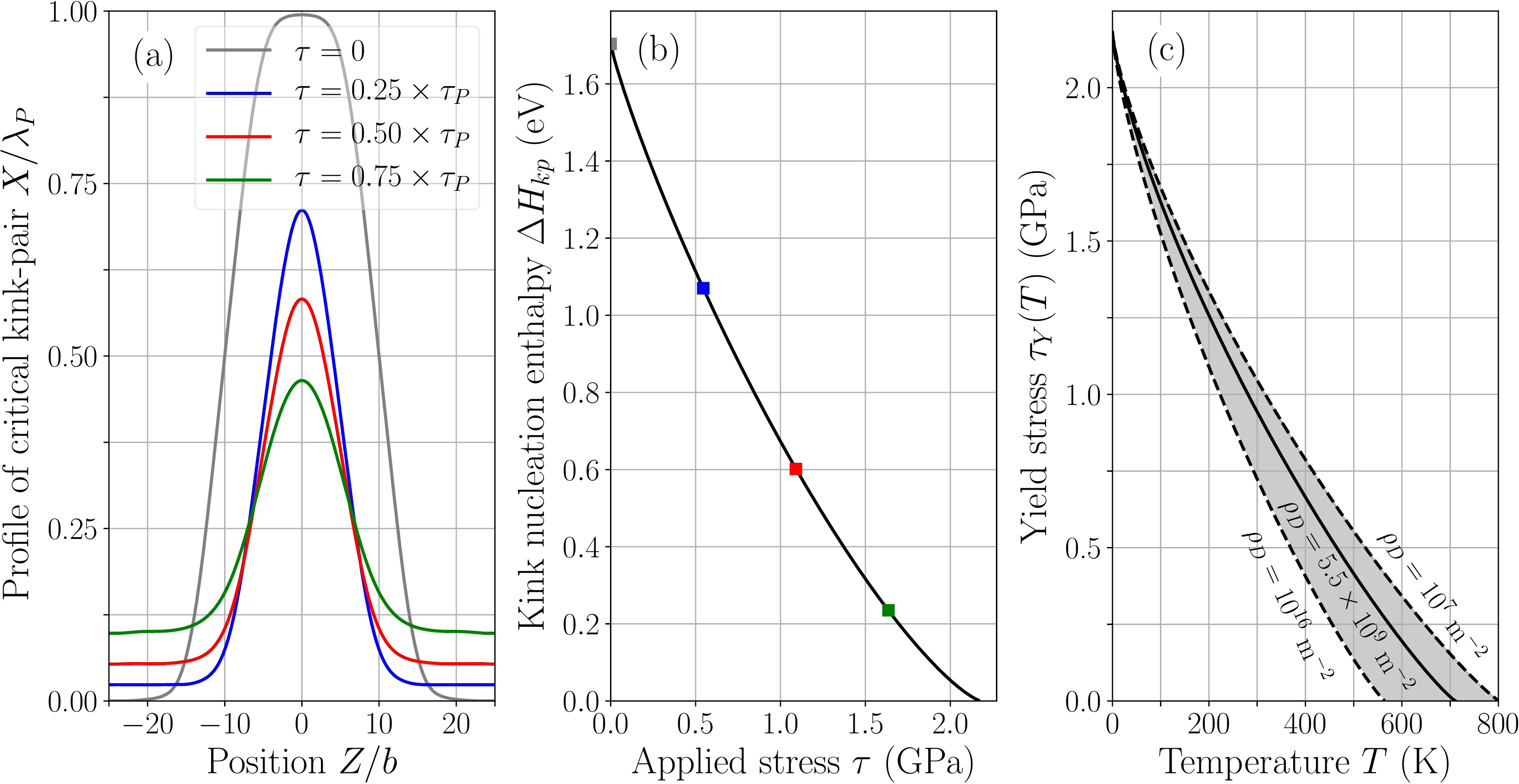}
	\end{center}
	\caption{Line tension model of dislocation mobility adjusted on \abinitio{} calculations:
	(a) line profile of a screw dislocation with a critical kink-pair for different applied resolved shear stresses,
	(b) kink-pair nucleation enthalpy as a function of applied resolved shear stress (the solid line is a fit to Kocks' law and the colored squares refer to the stresses in (a)),
	(c) yield stress as a function of temperature 
	with a strain rate $\dot{\varepsilon}=10^{-5}$\,s$^{-1}$,
	a dislocation length $l_{\rm D}=1/\sqrt{\rho_{\rm D}}$
	and dislocation densities ranging from $10^{7}$ to $10^{16}$\,m$^{-2}$.
	No activation entropy ($T_{\rm m}=\infty$ in Eq. \ref{eq:yield_temperature}) is considered for the yield stress. 
	}
	\label{fig:ThermalActivation}
\end{figure}

Having computed \abinitio{} the two material parameters entering the line tension model, 
\ie{} the Peierls enthalpy $\Delta H_{\rm P}^{\rm 1D}(x)$ and the line tension $\Gamma$, 
the kink-pair nucleation enthalpy is obtained by searching for the dislocation profile $x(z)$
crossing one Peierls valley, which corresponds to a saddle point of the functional 
in Eq. \ref{eq:LT}.
Profiles obtained for different amplitudes of an applied $\hkl<111>\hkl{110}$ resolved shear stress $\tau$ are shown in Fig. \ref{fig:ThermalActivation}(a), with the corresponding activation enthalpies $\Delta H_{\rm kp}$ 
in Fig. \ref{fig:ThermalActivation}(b). Here we used a simple Peierls potential without non-Schmid effects 
($\Delta \barT{\Omega}=0$ and $\bar{y}(x)=0$ in Eq. \ref{eq:Peierls1D_relax}).
The nucleation enthalpy can be fitted using Kocks' law \cite{Kocks1975}:
\begin{equation}
	\Delta H_{\rm kp}(\tau) = \Delta E_{\rm kp} \left[ 1 -  \left( \frac{\tau}{\tau^0_{\rm P}} \right)^p \right]^q,
	\label{eq:Kocks0}
\end{equation}
where $\Delta E_{\rm kp}$ is the kink-pair formation energy, 
$\tau^0_{\rm P}$ is the Peierls stress in the $\{110\}$ plane, 
and $p$ and $q$ are adjustable parameters. 
For BCC W, we obtain a formation energy $\Delta E_{\rm kp}=1.70$\,eV for two isolated kinks.
As demonstrated in Ref. \cite{Dezerald2015}, this formation enthalpy can be well approximated by
$\Delta E_{\rm kp} = 4\sqrt{2}/\pi \times \lambda_{\rm P}\sqrt{\Gamma \, V_{\rm P}^{\rm act}}$,
with $\lambda_{\rm P}=a\sqrt{6}/3$ the distance between Peierls valleys
and $V_{\rm P}^{\rm act}$ the height of the Peierls potential. 
This evidences that the kink formation enthalpy is equally sensitive to the Peierls potential
and the line tension.

Non-Schmid effects can be readily included in the line tension  approximation by using in Eq. \ref{eq:LT} 
the full expression of the Peierls enthalpy $\Delta H_{\rm P}^{\rm 1D}$ given by Eq. \ref{eq:Peierls1D_relax}.
The kink-pair nucleation enthalpy can still be described by Kocks' law, 
but with parameters that now depend on the mechanical loading. In the case of a uniaxial tensile test as discussed above, we have:
\begin{equation}
	\Delta H_{\rm kp}(\sigma,\zeta,\chi) = \Delta E_{\rm kp} \left[ 1 -  \left( \frac{\sigma}{\sigma^0_{\rm Y}(\zeta,\chi)} \right)^{p(\zeta,\chi)} \right]^{q(\zeta,\chi)}.
	\label{eq:Kocks}
\end{equation}

\subsection{Dislocation velocity and yield stress}

Assuming that dislocation glide is controlled by kink nucleation,
the dislocation velocity $v_{\rm gl}$ is given by:
\begin{equation}
	v_{\rm gl} = \nu_{\rm D} \dfrac{l_{\rm D}}{b} \lambda_{\rm P} 
	\exp{\left( -\dfrac{\Delta G_{\rm kp}}{k_{\rm B} T} \right)}. 
	\label{eq:velocity}
\end{equation}
$\nu_{\rm D}$ is an attempt frequency for the nucleation event
and is taken equal to the Debye frequency ($\nu_D=52$\,THz for BCC W \cite{Kittel1966}). The ratio $l_{\rm D}/b$ is an estimate of the number of potential kink-pair nucleation sites, with $l_{\rm D}$ the length of the dislocation line. A good estimate of this dislocation length is $l_{\rm D}=1/\sqrt{\rho_{\rm D}}$, with $\rho_{\rm D}$ the dislocation density.
The kink-pair formation free enthalpy 
$\Delta G_{\rm kp} =\Delta H_{\rm kp} - T \Delta S_{\rm kp}$ is composed of the formation enthalpy (Eq. \ref{eq:Kocks})
and of the formation entropy, which is unknown. 
The computational effort needed to evaluate this entropic contribution 
using either an harmonic approximation \cite{Proville2012} or thermodynamic integration \cite{Swinburne2018}
is still out of reach of \abinitio{} calculations.
Here, in order to evidence the potential impact of entropy, we will use a simple approximation, 
the Meyer-Neldel compensation rule \cite{Meyer1937}, which assumes that the activation entropy is proportional 
to the activation enthalpy, $\Delta S_{\rm kp} = \Delta H_{\rm kp}/T_{\rm m}$,
with the parameter $T_{\rm m}$ homogeneous to a temperature and expected to be close to the melting temperature ($T_{\rm m}=3695$\,K for W \cite{Kittel1966}).
We will first neglect entropic contributions by setting $T_{\rm m}=\infty$, before discussing their potential impact by choosing finite values for $T_{\rm m}$.
Altogether, the activation free enthalpy for kink-pair nucleation during a uniaxial tensile test is given by
\begin{equation}
	\Delta G_{\rm kp}(\sigma,\zeta,\chi,T) = \Delta E_{\rm kp}
	\left[ 1-\left( \dfrac{\sigma}{\sigma^0_{\rm Y}(\zeta,\chi)} \right)^{p(\zeta,\chi)} \right]^{q(\zeta,\chi)}
	\left( 1 - \dfrac{T}{T_{\rm m}} \right). 
	\label{eq:deltaG}
\end{equation}
Knowing the dislocation velocity, one can deduce the rate of plastic deformation $\dot{\varepsilon}$
for a given density of mobile dislocations $\rho_{\rm D}$
from Orowan's law, 
\begin{equation}
	\dot{\varepsilon} = \rho_{\rm D} \, b \, v_{\rm gl}.
	\label{eq:Orowan}
\end{equation}
Using Eq. \ref{eq:velocity} for the dislocation velocity and Eq. \ref{eq:deltaG} for the activation free enthalpy, Orowan's law can be inverted 
to obtain an expression of the yield stress $\sigma_Y$ of a given slip system during a tensile test at constant temperature and strain rate: 
\begin{equation}
	\sigma_{\rm Y}(T,\zeta,\chi) = \sigma^0_{\rm Y}(\zeta,\chi) 
	\ \left\{ 1 - \left[ \dfrac{k_{\rm B} T}{\Delta E_{\rm kp}} \dfrac{T_{\rm m}}{T-T_{\rm m}} 
	\ln{ \left( \dfrac{\dot{\varepsilon}}{\rho_{\rm D} \, \nu_{\rm D} \, l_{\rm D} \, \lambda_{\rm P} } \right) } 
	\right]^{1/q(\zeta,\chi)} \right\}^{1/p(\zeta,\chi)} .
	\label{eq:yield_temperature}
\end{equation}
Neglecting the entropic contribution ($T_{\rm m}=\infty$),
the critical temperature $T_{\rm c}$ at which the yield stress of \hkl<111>\hkl{110} slip systems vanishes is given by:
\begin{equation}
	T_{\rm c}^0=\frac{\Delta E_{\rm kp} }
	{ k_{\rm B} \ln{\left( {\rho_{\rm D} \, \nu_{\rm D} \, l_{\rm D} \, \lambda_{\rm P} }\,/\,
	{\dot{\varepsilon}}\right) }} .
	\label{eq:critical_temperature}
\end{equation}
We note that this critical temperature depends neither on the relative orientation of the tensile axis 
nor on the considered slip system.
It defines the athermal limit
above which the plastic deformation is no longer thermally activated. 
The variations in tungsten of the \hkl<111>\hkl{110} yield stress with the temperature and the resulting critical temperatures 
are presented in Fig. \ref{fig:ThermalActivation}(c), 
when non-Schmid effects are neglected. 
As illustrated in this figure, 
the critical temperature is sensitive to the dislocation density $\rho_{\rm D}$, 
both directly from Orowan's law (Eq. \ref{eq:Orowan})
but also indirectly through the dislocation length 
$l_{\rm D}=1/\sqrt{\rho_{\rm D}}$. 
It varies from about 560\,K to 800\,K when the dislocation density varies from very high ($10^{16}$ m$^{-2}$) to very low ($10^{7}$ m$^{-2}$) for a fixed strain rate $\dot{\varepsilon}=8.5\times10^{-4}$\,s$^{-1}$.

When entropy contributions are included (finite $T_m$), 
lower energy barriers for kink nucleation are obtained, allowing for easier dislocation glide at high temperatures and a lower critical temperature
$T_{\rm c} = T_{\rm c}^0 \,/\, (1 + T_{\rm c}^0/T_{\rm m})$.
For a dislocation density $\rho_{\rm D}=10^9$\,m$^{-2}$ and a strain rate $\dot{\varepsilon}=8.5\times10^{-4}$\,s$^{-1}$, the critical temperature goes from 732\,K without entropy to 611\,K when the Meyer-Neldel approximation is used with $T_{\rm m}=3695$\,K.

This sensitivity to the parameter $T_{\rm m}$ of the Meyer-Neldel approximation shows the impact of vibrational contributions to the yield stress at high temperatures. 
Accounting for the activation entropy associated with atomic vibrations in the kink pair nucleation free enthalpy
appears therefore necessary not only at low temperatures to resolve the discrepancy between  experimental and theoretical yield stresses \cite{Proville2012,Barvinschi2014,Proville2018b} (\cf{} section \ref{sec:PeierlsStress})
but also at high temperatures.  Moreover, anharmonicity may become important at high temperatures, requiring to use more precise approaches 
relying on thermodynamic integration \cite{Gilbert2013,Swinburne2018,Sato2021}. These approaches are still too computationally expensive for \abinitio{} calculations, particularly in the case of a kinked dislocation. 
Below we will neglect entropic contributions, 
thus using $T_{\rm m}=\infty$.

\subsection{Temperature-dependence of the yield stress in tungsten}

\begin{figure}[bht!]
	\begin{center}
		\includegraphics[width=1\linewidth]{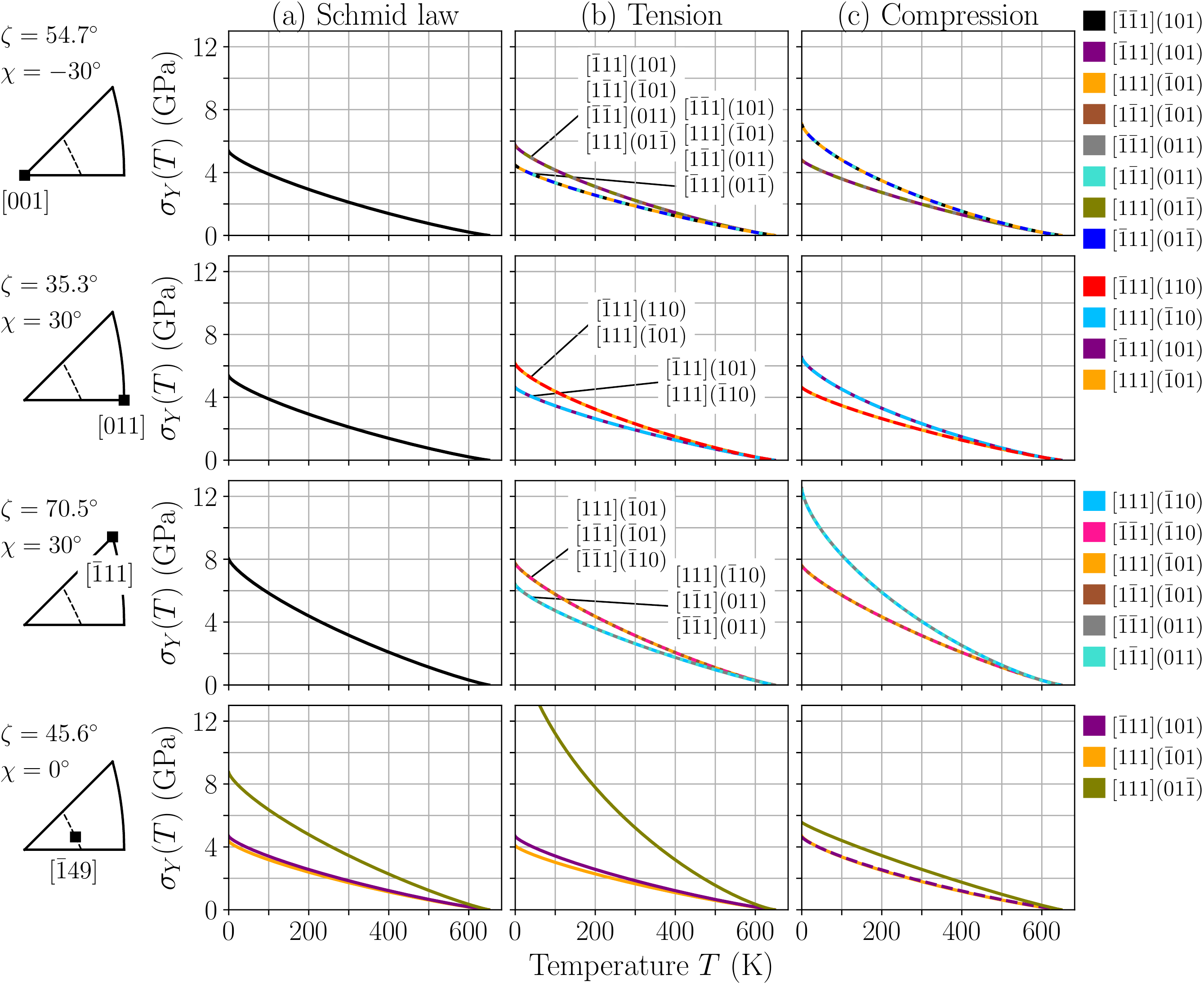}
	\end{center}
	\caption{Yield stress of the primary \hkl<111>\hkl{110} slip systems predicted from Eq. \ref{eq:yield_temperature} as a function of temperature for a tensile axis along \hkl[001] (first row), \hkl[011] (second row), \hkl[-111] (third row), and \hkl[-149] (last row), according to Schmid's law (a) and including non-Schmid effects in tension (b) and compression (c).
	The dislocation density is $\rho_{\rm D}=10^{12}$\,m$^{-2}$, 
	with $l_{\rm D}=1/\sqrt{\rho_{\rm D}}$, 
	a strain rate $\dot{\varepsilon}=8.5 \times 10^{-4}$\,s$^{-1}$,
	and no entropy contribution ($T_{\rm m}=\infty$). }
	\label{fig:TemperatureOrientations}
\end{figure}

\begin{figure}[bth]
	\begin{center}
		\includegraphics[width=1\linewidth]{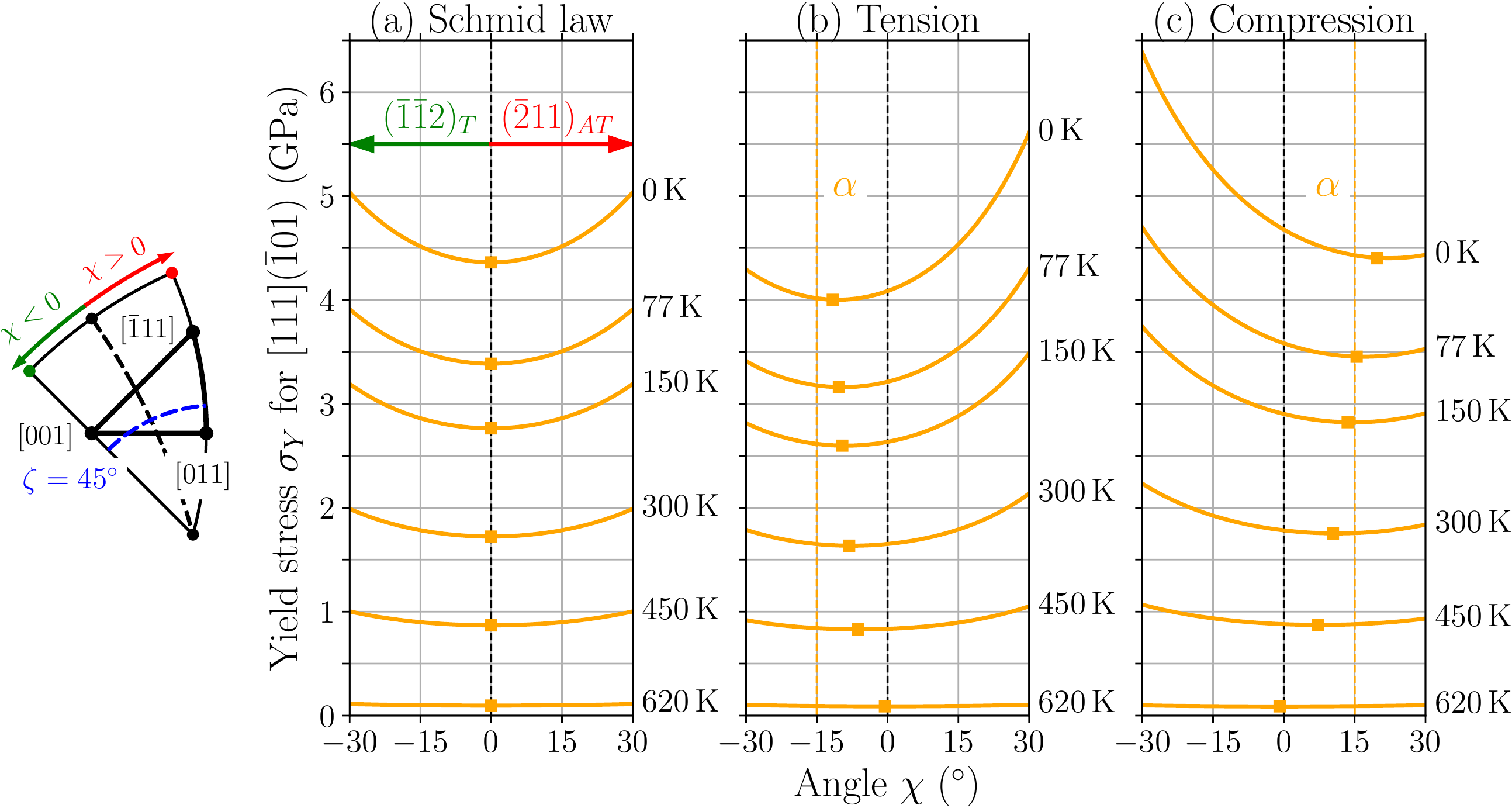}
	\end{center}
	\caption{Yield stress of the \hkl[111]\hkl(-101) slip system predicted from Eq. \ref{eq:yield_temperature} at different temperatures as a function of the angle $\chi$ for $\zeta=45^\circ$: according to Schmid's law (a) and including non-Schmid effects in tension (b) and compression (c). 
	The angle $\chi$ leading to a minimum yield stress is indicated by a square symbol.
	Parameters
	as in Fig. \ref{fig:TemperatureOrientations}
}
	\label{fig:ThetaLineTemperature}
\end{figure}

The model in Eq. \ref{eq:yield_temperature} is used in Fig. \ref{fig:TemperatureOrientations} to predict the temperature dependence of the yield stress of all slip systems for uniaxial tensile tests along the \hkl[001], \hkl[011] and \hkl[-111] corners of the standard stereographic triangle and along the more central \hkl[-149] orientation. For the corner orientations, the slip systems that are equivalent according to Schmid's law (Fig. \ref{fig:TemperatureOrientations}(a)) are split in two groups due to non-Schmid effects in tension (Fig. \ref{fig:TemperatureOrientations}(b)) and compression (Fig. \ref{fig:TemperatureOrientations}(c)), with the relative ease to activate one group reversed when the sign of the applied stress is reversed.

A notable feature is that the deviation from Schmid's law, both in terms of T/AT and tension/compression asymmetries become less pronounced with increasing temperature. This is better visualized in Fig. \ref{fig:ThetaLineTemperature} where the yield stress of the \hkl[111]\hkl(-101) system as a function of the angle $\chi$ at $\zeta=45^\circ$ is plotted at different temperatures ranging from 0\,K to the critical athermal temperature. 
At low temperatures, the deviations from Schmid's law are strong as reported at 0\,K in section \ref{sec:generalized_yield}. But with increasing temperature, the T/AT and tension/compression asymmetries become less pronounced and vanish close to the athermal temperature. This recovery of Schmid's law at high temperature has been reported experimentally in BCC transition metals \cite{Liu1972,Nawaz1975}, and was also accounted for using a model 2D Peierls potential coupled with a LT model in the work of Edagawa \textit{et al.} \cite{Edagawa1997}.

\begin{figure}[bth]
	\begin{center}
        \hspace*{2mm}
		\includegraphics[width=0.9\linewidth]{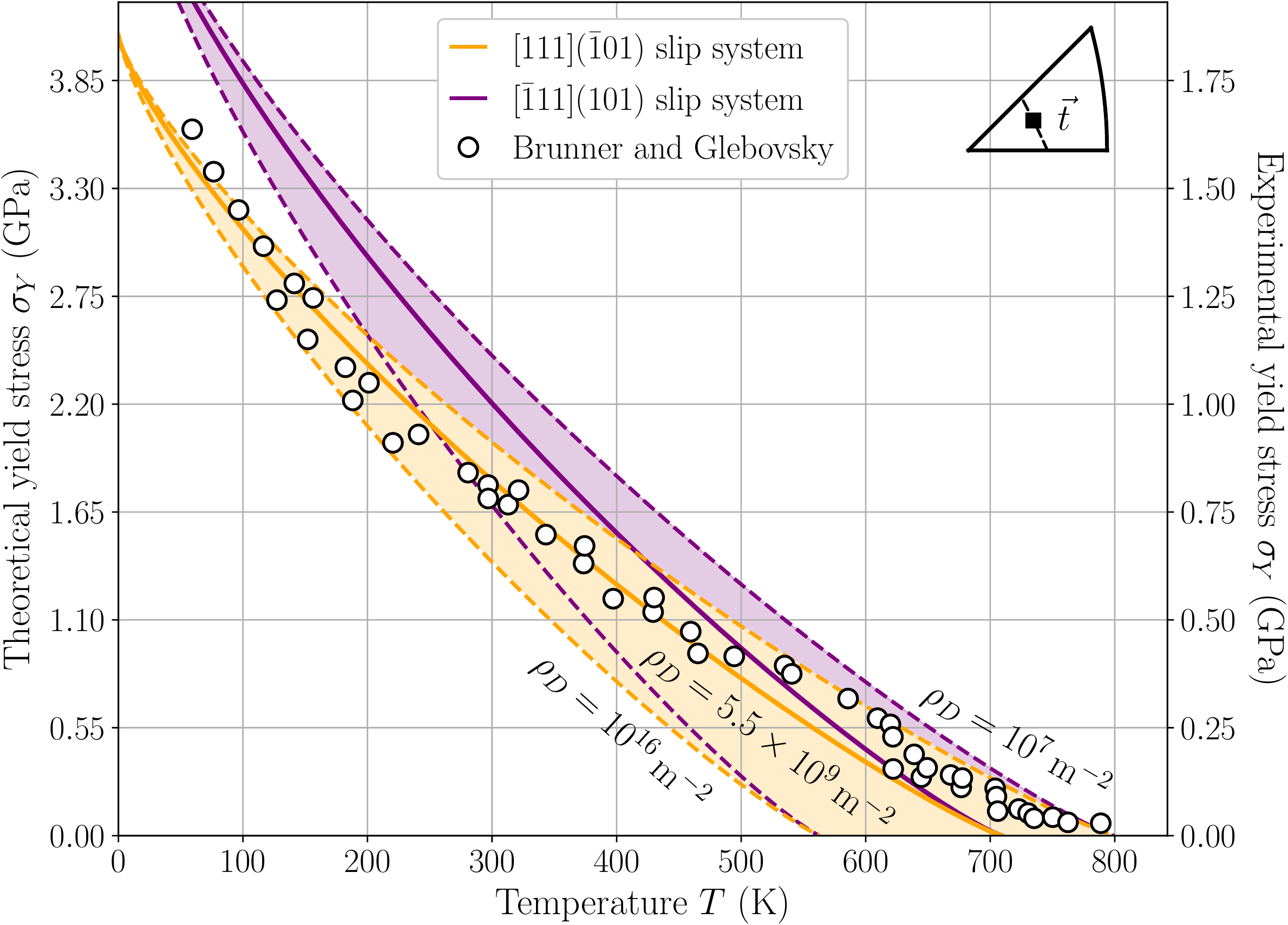}
	\end{center}
	\caption{Yield stress for \hkl[111]\hkl(-101) (orange) and \hkl[-111]\hkl(101) (purple) slip systems predicted from Eq. \ref{eq:yield_temperature} (left axis) and experimental data from Brunner and Glebovsky \cite{Brunner2000} (right axis). In both cases, the traction axis is along $\hkl[-149]$ ($\zeta=50^\circ$ and $\chi=0^\circ$) with a strain rate $\dot{\varepsilon}=8.5 \times 10^{-4}$\,s$^{-1}$. 
	The central bold lines correspond to the estimated experimental dislocation density $\rho_{\rm D}=5.5 \times 10^9$\,m$^{-2}$. 
	}
	\label{fig:BrunnerTemperature}
\end{figure}

We finally compare the predicted temperature dependence of the yield stress with experimental data from Brunner and Glebovsky \cite{Brunner2000} in Fig. \ref{fig:BrunnerTemperature} obtained in tension along an axis with $\zeta=50^\circ$ and $\chi=0$ close to $\hkl[-149]$. For this orientation, the generalized yield criterion in tension at 0\,K predicts that the \hkl[111]\hkl(-101) and \hkl[-111]\hkl(101) slip systems are the easiest to activate (see Fig. \ref{fig:ThetaLines}(b)). The variation of the yield stress for both slip systems is plotted as a function of temperature in Fig. \ref{fig:BrunnerTemperature} for different  dislocation densities $\rho_{\rm D}$ ranging from $10^{7}$ to $10^{16}$\,m$^{-2}$. 
In the experiments, the dislocation density is unknown. But in a later study on the temperature-dependent tensile properties of tungsten single crystals with a similar orientation close to \hkl[-149] ($\zeta=45.6^\circ$ and $\chi=0$) \cite{Brunner2010}, the authors reported a dislocation density of $5.5 \times 10^{9}$\,m$^{-2}$. Using this value as a reference, we highlighted the corresponding theoretical prediction in Fig. \ref{fig:BrunnerTemperature}. Note that as before, experimental data are shown with a different scale.

As expected, both theoretical and experimental yield stresses decrease with temperature and follow similar, slightly convex curves. Moreover, the predicted and experimental curves at the estimated experimental density are rather close, keeping in mind that the data are not shown with the same stress scale. Temperatures are however shown with the same scale and we see that the predicted athermal temperature, 710\,K, is close to the experimental value, about 800\,K. This is particularly satisfactory keeping in mind that we have neglected entropic effects and that the theoretical prediction does not use any fitting parameter. This agreement is an indication that  the kink-pair formation energy $\Delta E_{\rm kp}$, which is directly proportional to the critical temperature (Eq. \ref{eq:critical_temperature}), is correctly estimated by our modeling approach.

\section{Conclusions}

\Abinitio{} calculations reveal that many specific features of BCC metals plasticity 
can be rationalized by core properties of the $1/2\,\hkl<111>$ screw dislocation. 
The compact core of this dislocation results in elementary glide events in \hkl{110} planes,
but, because of the lack of inversion symmetry of these \hkl{110} planes, 
the trajectory of the gliding dislocation deviates from a straight path between stable positions.
This leads to the observed T/AT asymmetry of the yield criterion:
the shear stress necessary to activate plasticity is minimal 
when the MRSSP is tangent to the dislocation trajectory and not when it corresponds 
to the macroscopic \hkl{110} glide plane.
\Abinitio{} calculations also evidence that dislocations have a non-null relaxation volume  
which couples with non-glide components of the stress tensor.  
Variations of the relaxation volume along the dislocation glide path
lead to variations of the energy barrier opposing dislocation glide, and thus of the yield stress.
All these ingredients can be then incorporated in a line tension model to describe kink-pair nucleation 
and thus predict dislocation velocity as a function of temperature and mechanical loading,
either analytically using simple expressions based on classical nucleation theory like in the present work, or using kinetic Monte Carlo simulations \cite{Stukowski2015}.
Starting from \abinitio{} calculations, one thus obtains theoretically a full description of single crystal plastic yield below the critical athermal temperature where plasticity is controlled by the mobility of $1/2\,\hkl<111>$ screw dislocations.

Mobility laws for dislocation glide derived from this \abinitio{} description
can be directly implemented in dislocation dynamics simulations \cite{Po2016}
or crystal plasticity \cite{Cereceda2016}.
Such a multiscale approach allows accounting for collective effects,
with the dislocation mobility depending not only on the external mechanical loading 
but also on internal stresses \cite{Srivastava2020}.

Detailed comparison of \abinitio{} predictions with experimental data shows that this modeling approach
is still not fully quantitative. 
The most striking disagreement concerns the prediction of the Peierls stress at 0\,K,
with the theoretical value being two or three times larger than in experiments across BCC metals. 
To resolve, at least partly, the discrepancy, it appears necessary to consider not only 
the energy barrier opposing dislocation glide but also the associated activation entropy 
arising from atomic vibrations. 
Atomic simulations relying on empirical interatomic potential have shown 
that variations of the zero-point energy coming from the quantization of the vibrational modes 
is responsible for a lowering of the theoretical value of the Peierls stress at low temperatures \cite{Proville2012,Proville2018b}. 
Vibrations are also important at higher temperatures where they can significantly modify the Peierls potential \cite{Gilbert2013}
and the activation energy barrier for kink-pair nucleation \cite{Swinburne2018}. 
Determination of these entropy contributions with \abinitio{} calculations is still out of reach. 
It appears therefore necessary to rely on empirical potentials to obtain such entropy contributions.
Although simple EAM potentials have shown that they are versatile enough to reproduce key properties
of dislocations \cite{Mendelev2003,Marinica2013}, such central-force potentials do not contain 
the ingredients necessary to account for the angular dependence of interatomic bonding 
in BCC transition metals. 
More sophisticated descriptions of atomic interactions are therefore highly desirable 
to fill the gap between \abinitio{} calculations and simple empirical potentials:
semi-empirical approaches relying on tight binding approximations, 
like bond order potentials \cite{Mrovec2011}, 
or fully phenomenological approaches relying on machine learning, 
such as Gaussian approximation potentials \cite{Maresca2018} or neural network atomic potentials \cite{Mori2020}, are two promising possibilities.

A clear understanding of dislocation glide in \hkl{112} planes is also missing. 
Experiments \cite{Argon1966,Caillard2018} have unambiguously shown that dislocations with a $1/2\,\hkl<111>$ Burgers vector
can glide not only in \hkl{110} planes but also in \hkl{112} planes in some BCC metals,
for instance in tungsten at low temperature. 
No elementary glide mechanism compatible with the compact core structure of the $1/2\,\hkl<111>$ 
screw dislocation predicted by \abinitio{} calculations have been proposed until now 
to rationalize glide in \hkl{112} planes.

Magnetism is another challenge for the \abinitio{} modeling of dislocations. 
If ferromagnetic BCC metals like Fe do not introduce any supplementary technical difficulty 
compared to non-magnetic elements \cite{Ventelon2013,Proville2013,Dezerald2014,Dezerald2016},
the same is not true for paramagnetic and antiferromagnetic elements.
Fe and Cr are two BCC metals which become paramagnetic 
above respectively their Curie (1043\,K) and Néel (311\,K) temperatures.
Modeling of dislocations in these paramagnetic states requires to account for the disordering 
of the atom magnetic moments and to perform statistical averages on magnetic configurations. 
Such an \abinitio{} modeling approach in BCC Fe has shown that screw dislocations
have similar properties in the ferromagnetic and paramagnetic phases \cite{CasillasTrujillo2020}.  
On the other hand, magnetic order in an antiferromagnetic phase, like BCC Cr at low temperature\footnote{The 
real magnetic state of Cr at low temperature corresponds to a spin density wave
for which the antiferromagnetic phase is a good approximate.}, 
is not compatible with a $1/2\,\hkl<111>$ Burgers vector: 
dislocations with such a Burgers vector necessarily introduce a magnetic fault in the crystal.
It has been proposed that these dislocations coexist pairwise to bound the magnetic fault, 
thus leading to super-dislocations with a \hkl<111> Burgers vector
\cite{Bienvenu2020}.

The understanding of alloying effects on BCC plasticity
also benefits from the development of \abinitio{} calculations. 
Modeling of dislocation interactions with solute atoms sheds new light 
on the mechanisms responsible for hardening or softening in dilute solid solutions 
\cite{Trinkle2005,Itakura2013,Tsuru2020} 
and in more concentrated solid solutions like high entropy alloys
\cite{Yin2020}.
In addition, some substitutional solute atoms have been shown to induce a change of the core structure of the screw dislocation through the variation of the electronic density
\cite{Romaner2010,Li2012,Romaner2014,Samolyuk2013}, 
with the dislocation going from a symmetric compact to a degenerate polarized core.
\Abinitio{} calculations have also evidenced core reconstructions of the screw dislocation 
induced by interstitial solutes, 
with H stabilizing the split configuration \cite{Grigorev2020}
and bigger interstitial solute atoms like carbon stabilizing the hard core 
\cite{Ventelon2015,Luthi2017,Luthi2018,Luthi2019,Bakaev2019,Hachet2020}.
Integrating the elementary interaction mechanisms revealed by these \abinitio{} calculations
in higher scale models, allows to tackle more complex phenomena,
like the Portevin - Le Chatelier effect \cite{Zhao2020}
or the reappearance of a Peierls regime and dynamic strain ageing 
at temperatures where solute diffusion is activated \cite{Caillard2015,Caillard2016}.

\subsubsection*{Acknowledgments}
Antoine Kraych, Lisa Ventelon, and François Willaime are acknowledged
for their contributions to the works presented here.
Part of this work has been performed using HPC resources from GENCI-CINES and -TGCC 
(Grants 2020-096847 and -0910156). 
The authors also acknowledge PRACE for access to Juwels system hosted by Jülich Supercomputing Centre, Germany (project DIMAB).
LD acknowledges support from the French State through the program “Investment in the future” operated by the National Research Agency (ANR) and referenced by ANR-11-LABX-0008-01 (LabEx DAMAS)

\appendix

\section{\Abinitio{} parameters and dislocation setup}
\label{app:abinitio}

All previously unpublished results presented in this work were obtained using the \textsc{Vasp} code. The electronic structure of tungsten is modeled using a projector augmented wave pseudopotential with 14 valence electrons, and the GGA-PBE functional is used to approximate the exchange and correlation potential. An energy cutoff of 600\,eV is chosen for the plane-wave basis and the Brillouin zone is sampled using a $\Gamma$-centered $k$-point grid generated with the Monkhorst-Pack scheme and 24 $k$-points per lattice parameter unit length in every direction. Structural relaxations are performed with fixed periodicity vectors and a convergence criterion imposing that forces in all Cartesian directions are less than 5\,meV/{\AA}. The nudged elastic band (NEB) method as implemented in the \textsc{Vasp} code is used to find the minimum energy path between atomic configurations with 5 intermediate images and a spring constant of 5\,eV/{\AA}.

The dislocation dipole is introduced in a supercell containing 135 atoms
with periodicity vectors 
$\vec{U}_1=5/2\,\hkl[-12-1]+9/2\,\hkl[-101]$, 
$\vec{U}_2=5/2\,\hkl[-12-1]-9/2\,\hkl[-101]$, 
and $\vec{U}_3=1/2\,\hkl[111]$.
At variance with our previous works \cite{Dezerald2014,Dezerald2016,Kraych2019}, 
the quadrupolar positions of the dislocation periodic array,
for which the vector separating two dipole dislocations is  closest to 
$(\vec{U}_1+\vec{U}_2)/2$,
is chosen to coincide with the saddle point of the Peierls barrier. 
This choice minimizes the variation of elastic energy along the path
and respects the inversion symmetry of this path, 
ensuring in particular that the initial and final configurations 
are fully equivalent.

\section{MRSSP and resolved shear stress}
\label{app:chi_exp}

Experimental data for the variations of the yield stress 
with the orientation of the tensile axis 
\cite{Rose1962,Beardmore1965,Argon1966}
are expressed with angles $\phi$ and $\psi$
defining the orientation of the tensile axis $\vec{t}$ with respect to the corners of the 
\hkl[001], \hkl[011] and \hkl[-111] stereographic triangle 
(see inset in Fig. \ref{fig:EdgesTriangle_b111} for a definition of these angles).
We give here the correspondence with the angle $\chi$ defining the MRSSP (Fig. \ref{fig:trajectory})
and the corresponding resolved shear stress $\tau$ in this plane.

\begin{itemize}
	\item \hkl[001]-\hkl[011] edge: 
		\begin{equation*}
			\vec{t} = \left[ 0, \sin{(\phi)}, \cos{(\phi)} \right]
			\textrm{ with } 
			0\leq\phi\leq\pi/4
		\end{equation*}
		\begin{equation*}
			\chi(\phi) = \arctan{\left[ \frac{2 \tan{(\phi)} - 1}{\sqrt{3}} \right]}
			\textrm{ and }
			\tau(\sigma,\phi) = \frac{\sqrt{2}}{3} 
			\frac{ \cos^3{(\phi)} + \sin^3{(\phi)} }{ \sqrt{ 1 - \cos{(\phi)} \sin{(\phi)}}} \sigma
		\end{equation*}
	\item \hkl[011]-\hkl[-111] edge:
		\begin{equation*}
			\vec{t} = \left[ -\sin{(\psi)}, \frac{\cos{(\psi)}}{\sqrt{2}}, \frac{\cos{(\psi)}}{\sqrt{2}}  \right]
			\textrm{ with }
			0 \leq \psi \leq \arctan{\left( \frac{\sqrt{2}}{2} \right)}
		\end{equation*}
		\begin{equation*}
			\chi(\psi) = \frac{\pi}{3}
			\textrm{ and }
			\tau(\sigma, \psi) =  \frac{1}{6} \left( 
			2\sqrt{2}\cos{(2\psi)} + \sin{(2\psi)} \right) \sigma
		\end{equation*}
	\item \hkl[-111]-\hkl[001] edge:
		\begin{equation*}
			\vec{t} = \left[ -\frac{\cos{(\psi)}}{\sqrt{2}}, 
				\frac{\cos{(\psi)}}{\sqrt{2}},
				\sin{(\psi)} \right]
			\textrm{ with }
			\arctan{\left( \frac{\sqrt{2}}{2} \right)} \leq \psi \leq \pi/2
		\end{equation*}
		\begin{equation*}
			\chi(\psi) = \arctan{\left[ \frac{\sqrt{3}}{3}
			\frac{ 3\cos{(\psi)} - \sqrt{2}\sin{(\psi)} }
			{ \cos{(\psi)} + \sqrt{2}\sin{(\psi)} } \right]}
			\textrm{ and }
			\tau(\sigma,\psi) = \frac{\sqrt{2}}{6} \sin{(\psi)} \sqrt{5 + \cos{(2\psi)}} \, \sigma
		\end{equation*}
\end{itemize}

\section{Elastic constants of the deformed cell}
\label{app:elastic}

In section \ref{sec:relaxation_volume}, the elastic constants used to extract the relaxation volume tensor and the dislocation trajectory from the stress variations along the minimum energy path between Peierls valleys were adjusted from the bulk values to ensure the following conditions:
\begin{equation}
	x(\xi=0)=0 \ ;\qquad x(\xi=1)=\lambda_P \ ;\qquad  \Delta \Omega_{12}(\xi=0)=\Delta \Omega_{12}(\xi=1/2)=\Delta \Omega_{12}(\xi=1)=0 .
	\label{eq:Cij_fit_constraint}
\end{equation}
We evaluate here the elastic constants of a 135-atoms supercell, either perfect or containing a dislocation dipole, in order to validate the adjustment of the elastic constants performed in the main text.
The purpose is to show that this fitting of the elastic constants is compatible 
with the variations of the elastic constants induced by the dislocation because of anharmonicity.

In the frame of the gliding dislocation (\cf{} axis in Fig. \ref{fig:PBC_sketch}), the stress $\sigma_{ij}$ resulting from a deformation $\varepsilon_{ij}$ is given by:
\begin{equation*}
\begin{bmatrix}
\sigma_{11} \\ \sigma_{22} \\ \sigma_{33} \\ \sigma_{23} \\ \sigma_{13} \\ \sigma_{12}
\end{bmatrix}
=\begin{bmatrix} 
C_{11} & C_{12} & C_{13} & 0 & C_{15} & 0 \\
C_{12} & C_{11} & C_{13} & 0 & -C_{15} & 0 \\
C_{13} & C_{13} & C_{33} & 0 & 0 & 0 \\
0 & 0 & 0 & C_{44} & 0 & -C_{15} \\
C_{15} & -C_{15} & 0 & 0 & C_{44} & 0 \\
0 & 0 & 0 & -C_{15} & 0 & C_{66}
\end{bmatrix}.
\begin{bmatrix}
\varepsilon_{11} \\ \varepsilon_{22} \\ \varepsilon_{33} \\ 2\varepsilon_{23} \\ 2\varepsilon_{13} \\ 2\varepsilon_{12}
\end{bmatrix}
\end{equation*}
Deformations $\varepsilon^n$ were imposed to the periodicity vectors $\vec{U}_i$ of the cell according to:
\begin{equation*}
\vec{U}_i^n=(\mathbb{I}+\varepsilon^n) \times \vec{U}_i,\,\text{with}\,\varepsilon^n \in \varepsilon[1\,0\,0\,0\,0\,0],\ \varepsilon[0\,1\,0\,0\,0\,0],\ \ldots
\end{equation*}
To evaluate the elastic constants of the supercells, we minimize the following cost function $\mathcal{C}$ of the elastic tensor $C$ over the deformations $\varepsilon^n$ and corresponding stresses $\sigma^n$:
\begin{equation*}
	\mathcal{C}(C_{IJ})=\sum_{n} \bigg[ \sum_{I} \left( \sigma_{I}^n -C_{IJ} \varepsilon_{J}^n \right)^2 \bigg],
	\label{eq:cost_function}
\end{equation*}
with $I$ and $J$ indexes of tensors in Voigt notation.

\begin{table}[htb]
	\caption{Elastic constants $C_{IJ}$ (in GPa) in the frame of the $1/2\,\hkl<111>$ screw dislocation 
	(\cf{} axis in Fig. \ref{fig:PBC_sketch})
	obtained with the 2-atom perfect cubic cell (first row), 
	or with the 135-atom supercell considering either the perfect (second row) or the dislocated (third row) crystal.
	The last row gives the elastic constants obtained through a fit to enforce the constraints given by Eq. \ref{eq:Cij_fit_constraint}.
	Only values in italic were allowed to vary in the fit.
	}
	\label{tab:elastic_corr}
	\centering
	\begin{tabular}{lr c c c c c c c}
		\hline\hline
		 && $C_{11}$ & $C_{12}$ & $C_{13}$ & $C_{15}$ & $C_{33}$ & $C_{44}$ & $C_{66}$ \\
		\hline
		Perfect crystal		&   2-atoms cell & 495.3 & 206.9 & 210.5 & \textit{5.1} & 491.7 & \textit{147.9} & 144.2 \\
					& 135-atoms cell & 495.5 & 202.9 & 213.3 &        {7.3} & 486.7 &        {152.3} & 145.8 \\
		Dislocated crystal 	& 135-atoms cell & 498.6 & 213.1 & 209.5 &        {5.4} & 486.5 &        {137.4} & 142.5 \\
		Fit			& 135-atoms cell & 495.3 & 206.9 & 210.5 & \textit{4.5} & 491.7 & \textit{141.1} & 144.2 \\
		\hline\hline
	\end{tabular}
\end{table}

The obtained elastic constants are presented in Table \ref{tab:elastic_corr}
for both the perfect and the dislocated crystal using the 135-atoms supercell, 
along with the elastic constants in the perfect 2-atom cubic BCC unit cell.
The modified values  used in section \ref{sec:relaxation_volume} to correct the dislocation trajectory and relaxation volume tensor are reported in the last row. 
This table shows that the modifications of the elastic constants needed to enforce the conditions in Eq. \ref{eq:Cij_fit_constraint} 
is of the same order as the variations of these elastic constants induced by the presence of the dislocation dipole in the supercell. 
Comparing the values obtained for the perfect BCC crystal in the 2- and 135-atom cells, one also sees that this variation
is comparable to the precision expected for \abinitio{} calculations of elastic constants.

\bibliographystyle{crunsrt}
\bibliography{clouet2021}

\end{document}